\documentclass[journal=esthag,manuscript=article,layout=twocolumn]{achemso}
\usepackage[version=3]{mhchem} 
\usepackage{rotating}
\usepackage{pdflscape}
\usepackage{longtable}
\usepackage{hyperref}
\usepackage{xspace} 
\usepackage{xcolor}
\usepackage{amsmath}
\usepackage{amssymb}
\usepackage{tabularx}
\usepackage{multirow}
\usepackage{booktabs}
\newcommand{\K}{\,K\xspace} 
\newcommand{\ra}[3]{#1$^{\mathrm h}$#2$^{\mathrm m}$#3$^{\mathrm s}$} 
\newcommand{\dec}[3]{#1$^\circ$#2$^\prime$#3$^{\prime\prime}$} 

\author{Víctor M. Rivilla}\email{vrivilla@cab.inta-csic.es}
\affiliation{Centro de Astrobiología (CAB), CSIC-INTA, Ctra. de Ajalvir, km. 4, Torrejón de Ardoz, E-28850 Madrid, Spain}
\author{Miguel Sanz-Novo}
\affiliation{Centro de Astrobiología (CAB), CSIC-INTA, Ctra. de Ajalvir, km. 4, Torrejón de Ardoz, E-28850 Madrid, Spain}
\author{David San Andrés}
\affiliation{Centro de Astrobiología (CAB), CSIC-INTA, Ctra. de Ajalvir, km. 4, Torrejón de Ardoz, E-28850 Madrid, Spain}

\title[Interstellar stereoisomerism]
  {Interstellar stereoisomerism}

\abbreviations{IR,NMR,UV}

\begin{document}

\begin{tocentry}





\hspace{3mm}
\includegraphics[width=7.5cm,angle=0]{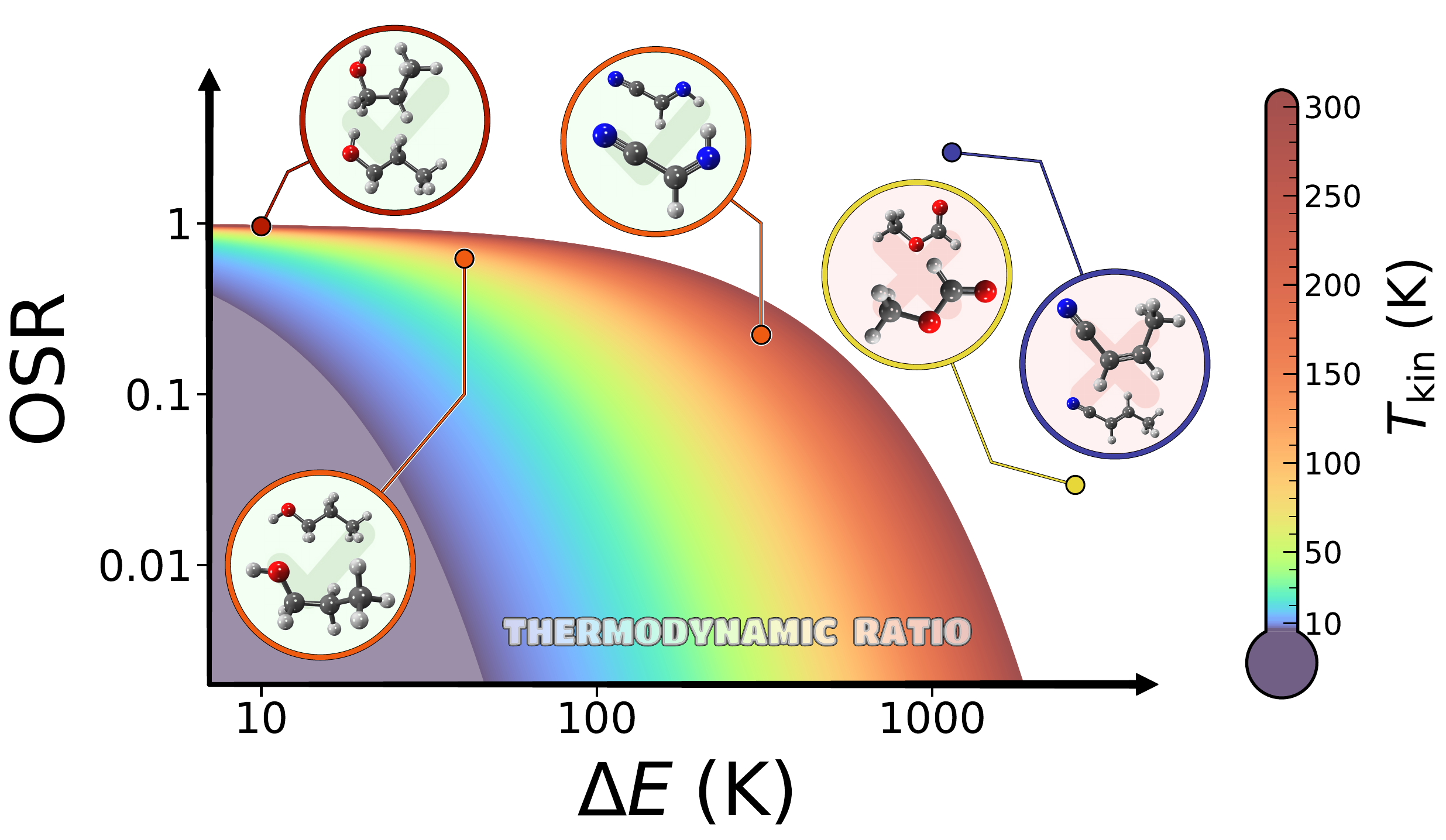}

\end{tocentry}

\begin{abstract}

The increasing detection of new molecules in the interstellar medium (ISM) shows that stereoisomerism is a fundamental contributor to interstellar molecular complexity. 
This work presents the first comprehensive overview of interstellar stereoisomerism.
%
A total of 16 stereoisomeric pairs have been identified (13 conformational and 3 geometric), spanning molecules with 5–12 atoms and energy separations from $\sim$10 K to 2667\K. They were observed across diverse astrophysical environments with kinetic temperatures ranging from low to high values ($\sim$7.5 K to 300 K).
%
The observed stereoisomeric ratios (OSR) - defined as the column density ratio of the higher-energy isomer divided by that of the lower-energy isomer - vary widely (0.009$–$4).
%
%
While systems with small energy differences ($<$ 600 K; i.e., $\sim$1.2 kcal mol$^{-1}$) in hot environments ($>$ 100 K) generally follow thermodynamic expectations (often assisted by tunneling-driven interconversion), many stereoisomers — particularly those in cold clouds or with larger energy separations— exhibit abundances far exceeding equilibrium values. This demonstrates that thermodynamics alone cannot explain interstellar stereoisomerism. Instead, stereoselective formation/destruction pathways (in the gas phase and/or in the surface of dust grains), 
photoisomerization, and chemical rearrangement during desorption must play a dominant role.
Stereoisomeric ratios thus provide powerful constraints on interstellar chemical pathways, and about the physico/chemical conditions of the ISM. This review highlights the need for stereochemistry-sensitive astrochemical models. Progress in this field requires expanded laboratory spectroscopy of higher-energy stereoisomers, dedicated quantum chemical studies of isomerization processes, and the explicit inclusion of stereoselective chemistry in chemical networks. Together, these efforts will be essential for understanding the origin of stereoisomeric selectivity and molecular complexity in the ISM.

\end{abstract}

\section{\rm \bf \large $\blacksquare$  INTRODUCTION}
\label{sec-intro}

\begin{figure*}
\begin{center}
\includegraphics[width=17cm]{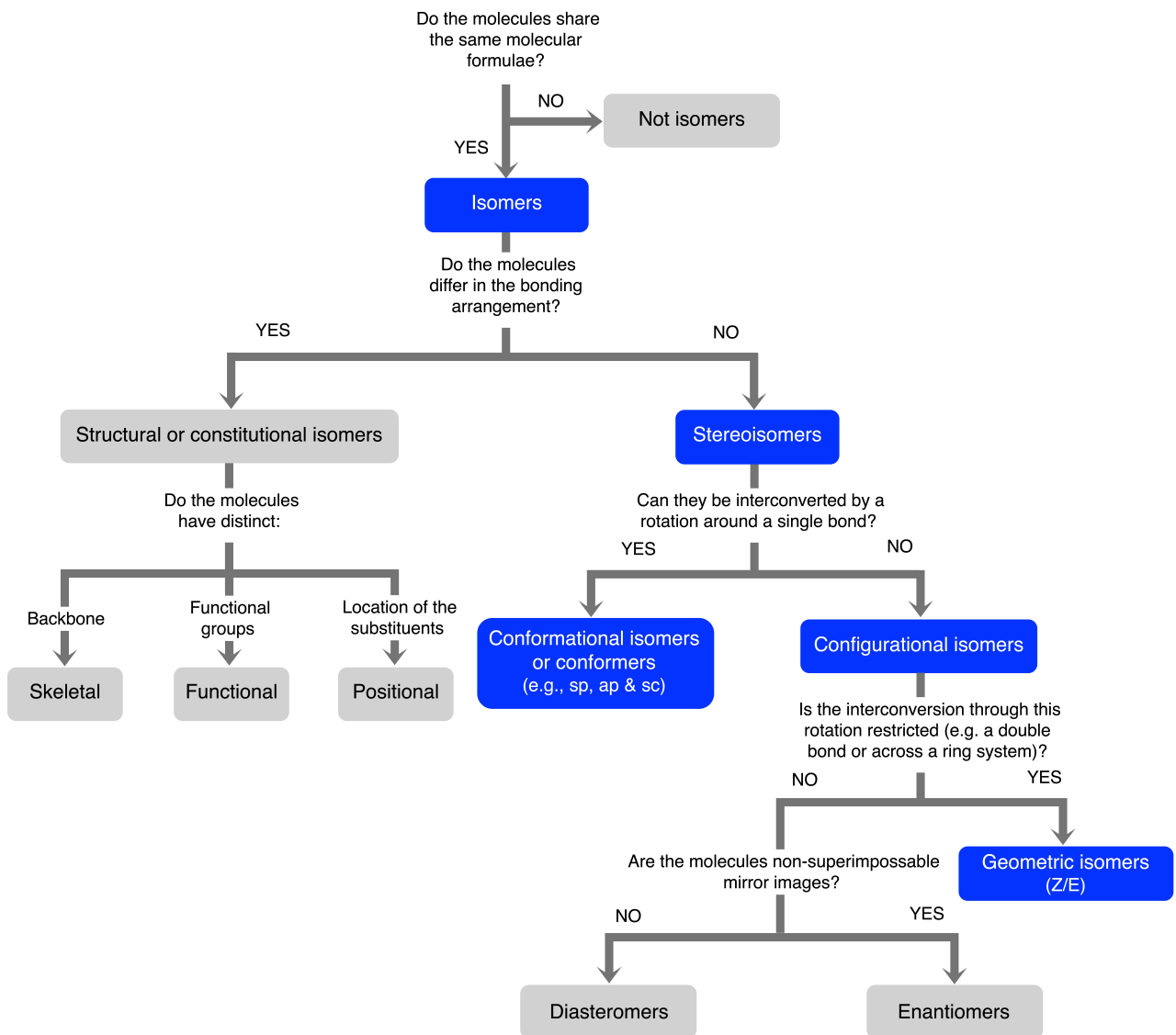}
\end{center}
\caption{Schematic view of the different types of isomerism. We highlight in blue the cases of stereoisomerism detected in the ISM, which are the topic of this work.}
\label{fig-diagram}
\end{figure*}

Since the pioneering very first detections of molecules in the interstellar medium (ISM) in the late 1930s\cite{Swings1937}, the census of interstellar species is continuously growing ($>$340 to date; see CDMS website\footnote{https://cdms.astro.uni-koeln.de/classic/molecules}), with a spectacular increase in the detection rate in the last few years \cite{McGuire2022}. 
A particular interesting group of interstellar molecules is composed by isomers, namely, those species with the same molecular formulae but different arrangement of atoms. 
%
%
There are different kinds of isomers,
which are described in Figure \ref{fig-diagram}. On one side, we find structural (or constitutional) isomers, which differ in their bonding arrangements. 
An example of widely-studied family of structural isomers comprises the species with chemical formula \ce{C2H4O2}, for which four different isomers have been identified:
methyl formate (\ce{CH3OCHO} \cite{Brown1975}), acetic acid (\ce{CH3COOH} \cite{Mehringer1997}), glycolaldehyde (\ce{HCOCH2OH} \cite{Hollis2000}), and 1,2-ethenediol (\ce{(CHOH)2} \cite{Rivilla2022}).

On the other side, isomers with the same bonding arrangement are known as stereoisomers, and they can be classified in two main groups (Figure \ref{fig-diagram}): i) conformational isomers (or conformers), if they can be interconverted by a rotation around a single bond; and configurational isomers, if not. Among the latter, if they differ in the spatial arrangement around a bond with restricted rotation (e.g. a double bond or across a ring system), they are classified as geometric isomers. If not, they are known as enantiomers or diasteromers, depending on whether or not they are superimpossable mirror
images. 
Hereafter, the term stereoisomer is used throughout this review to refer exclusively to those that have been detected in the ISM, which are conformational and geometric isomers
\footnote{We note that enantiomers cannot be distinguished using conventional rotational spectroscopic techniques because alternative approaches such as chiral tagging or three-way mixing measurements \citep{Patterson2013,Mills2022} are currently inaccessible in space.}.

Traditionally, the relative energies between isomers have been employed as a rule of thumb to evaluate their likelihood of detection in the ISM. \citet{Lattelais2009} proposed that, within a given molecular formula, the most abundant species should correspond to the most thermodynamically stable one, an empirical rule known as the Minimum Energy Principle (MEP). 
The authors analyzed 14 (structural) isomeric families detected in different astronomical sources, and found that the isomeric ratio might be qualitatively explained according to the energy difference between isomers, i.e., it is governed by thermodynamics.

Although \citet{Lattelais2009} already noted that the MEP was not fulfilled for some isomeric families, such as the \ce{C2H4O2} isomers mentioned above, the astrochemical community generally accepted the MEP as a valid rule.
However, the detections of new structural isomers over the last decade have revealed the failure of MEP in several other cases, such as in the \ce{C3H2O}, \ce{C2H4O2}, \ce{C2H2N2} or \ce{C2H5O2N} isomeric families \citep{Shingledecker2019,Mininni2020,Rivilla2023,SanAndres2024}, challenging its universality. This evidence points towards the role of alternative mechanisms, including chemical kinetics, i.e., different formation/destruction routes for each structural isomer, as the origin of the observed isomeric ratios.



In the case of stereoisomers, the prevailing view that MEP holds, in combination with the fact that the isomerization energy barriers from the lower-energy to the higher-energy stereoisomer ($E_{\rm iso}$) are usually too high to allow a direct interconversion at the low temperatures of the ISM, have contributed to perpetuate the idea that higher-energy stereoisomers should not be present in the ISM.
However, it is becoming evident, as a consequence of the increasing number of interstellar detections of higher-energy stereoisomers (e.g., \cite{Neill2012,Cuadrado2016,Rivilla2019,Sanz-Novo2023,Sanz-Novo2025,Zeng2025}), that this question requires further consideration. Indeed, the current number of stereoisomeric pairs identified in the ISM starts to allow a dedicated evaluation of their observed isomeric ratios, and thus to shed light about its possible origin. 

While the MEP appears to remain still valid for various stereoisomeric pairs, 
such as the conformers of ethanol (\ce{CH3CH2OH} \cite{Pearson1997}),
ethyl formate (\ce{CH3CH2OCHO}  \cite{Tercero2013,Rivilla2017}),
$n$-propanol (\ce{CH3CH2CH2OH}  \cite{Jimenez-Serra2022})
 or the $Z$/$E$ isomers of cyanomethanimine (\ce{HNCHCN} \cite{Rivilla2019,SanAndres2024}), there are also several counterexamples 
whose abundance ratios does not follow the thermodynamic expectation, including the 
conformers of formic acid (HCOOH \cite{GarciadelaConcepcion2022}), 
carbonic acid (HOCOOH \cite{Sanz-Novo2023}), or 
methyl formate  (\ce{CH3OCHO} \cite{Neill2012,Faure2014,Sanz-Novo2025}). 
Despite the increasing role of stereoisomers in new interstellar detections, the study of their chemistry remains markedly underexplored. Most theoretical and experimental studies focus only on constitutional isomers,
and do not distinguish between stereoisomers. Indeed, the chemical networks commonly used in astrochemical models, such as UMIST\footnote{https://umistdatabase.uk/} \citep{Millar2022} or KIDA\footnote{https://kida.astrochem-tools.org/} \citep{Wakelam2012}, are in the vast majority of cases insensitive to stereoisomerism, and consequently they are unable to explain the observed stereoisomeric ratios. 
Only a small number of theoretical works have addressed the study of interstellar stereoisomerism, and most of these were published in recent years (e.g., \cite{Neill2011,GarciadelaConcepcion2021,Molpeceres2021,Sanz-Novo2025,Molpeceres2025,Mallo2025}).

However, stereoisomeric ratios can provide useful constraints about the underlying  chemistry. 
Without understanding how the population of the species are distributed in the different stereoisomers, our knowledge about the mechanisms responsible for forming, destroying and isomerizing them will be very limited.

This work aims to fill that gap by providing the very first systematic overview of interstellar stereoisomerism. 
The manuscript is organized as follows. First, we describe and propose the general use of the updated nomenclature of stereoisomers based on modern IUPAC\footnote{International Union of Pure and Applied Chemistry (IUPAC); https://iupac.org/} recommendations, and describe the theoretical calculations used to compute the structures and energetics of the stereoisomers studied.
Afterwards, we
compile all of the stereoisomeric pairs detected in the ISM (including conformational and geometric), and
quantify their observed stereoisomeric ratios.
Subsequently, we 
discuss the different mechanisms that can contribute to shape these ratios in the ISM.
Finally, the concluding remarks arising from this review are presented.
Our
results underscore the need to fully incorporate stereochemistry into future astrochemical studies, which will be essential for guiding dedicated observational, laboratory, theoretical and modelling efforts aimed at uncovering the physical and chemical processes that rule stereoisomerism in the ISM.

\section{\rm \bf \large $\blacksquare$  RECOMMENDED NOMECLATURE FOR STEREOISOMERS}
\label{sec-nomenclature}

Regarding the nomenclature used to refer to different stereoisomers, we note that there are some discrepancies in the literature, which might cause some confusion. 
In an attempt to clarify this issue, we propose that, from now on, the astrochemical community uses the modern terminology approved by the IUPAC, which is the worldwide-recognized authority on chemical nomenclature, and recommends unambiguous, uniform, and consistent terminology. In particular, the {\it IUPAC Compendium of Chemical Terminology}\citep{IUPACGoldBook}, informally known as the ``Gold Book'', is the authoritative resource in this regard. We briefly discuss here the different terms used in previous astrochemically-related literature to refer to stereoisomers, and propose the use of the IUPAC recommended nomenclature.

For conformational isomers (conformers), the terms $gauche$, $anti$ and $syn$  have been commonly used. They refer to spatial relationships that arise from rotation around a single bond (i.e., varying a single dihedral angle), and therefore, correspond to conformations that interconvert freely without breaking any bond. We note, however, that in modern IUPAC terms,\citep{IUPACGoldBook} conformers traditionally described as $gauche$ 
correspond to $synclinal$ (sc), with dihedral angles around $\pm$60°, while those labeled $anti$ 
correspond to $antiperiplanar$ (ap), with dihedral angles near 180°. Conformers previously called $syn$ should be denoted as $synperiplanar$ (sp), with dihedral angles near 0°.\citep{Motiyenko2010} 



The use of $cis$/$trans$ or $E$/$Z$ notation for conformers, although it also frequently appears in the literature, is conceptually incorrect. These notations were traditionally applied to geometric isomers whose relative arrangement is fixed by restricted rotation (e.g., in alkenes or cyclic systems). Particularly, $cis$ and $trans$ describe the relative disposition of the substituents, which are located either on the same or opposite faces of a rigid framework. These notations were subsequently replaced by $E$ and $Z$, defined by the Cahn–Ingold–Prelog priority rules\cite{Cahn1996,Cahn1996b}, which designate the relative orientation of the highest-priority substituents across a double bond. These latter $E$ and $Z$ terms should therefore be used only for geometric isomers (see Figure \ref{fig-diagram}) rather than to describe conformational orientations that are inherently dynamic.
%

Hereafter in this review, we will always use the IUPAC recommended nomenclature, and encourage the astrochemical community to follow this criteria, although in some cases we will also make reference to old descriptors used in previous bibliography to allow an easy correspondence.





\section{\rm \bf \large $\blacksquare$  THEORETICAL CALCULATIONS}
\label{sec-calculations}

As mentioned before, stereoisomers share the same bonding arrangement, but they differ in their 3-dimensional (3D) stuctures in space. To illustrate the 3-dimensional representations of the stereoisomers detected in the ISM (described in next section), we compute the molecular structures in this work using the B3LYP hybrid density functional \citep{Becke_1993,LYP_1988} along with the aug-cc-pVTZ basis set \citep{cc-pVXZ_1989,Kendall1992,Woon1993}. We also consider the D3 version of Grimme’s dispersion to account for long range interactions.\citep{Grimme2010} We employ the Gaussian 16 program package to carry out the theoretical calculations,\cite{Frisch2016} and use IQmol\footnote{\url{https://www.iqmol.org/}} to visualize the structures. We note that after each geometry optimization, we performed a calculation of harmonic vibrational frequencies at the same level of theory used to optimize the geometry, to characterize the stationary points (i.e., to adequately identify real minima if they have all real vibrational frequencies), and to determine the zero-point vibrational energy (ZPE).

Moreover, for the further discussion in the next sections about the stereoisomeric values, we also compute the 
isomerisation energy barrier ($E_{\rm iso}$) for those interstellar stereoisomers for which these values were not previously reported in the literature. We determine them here from relaxed potential energy surface scans obtained by varying the corresponding dihedral angle at the B3LYP-GD3/aug-cc-pVTZ level of theory (in changes of 5$^{\circ}$, 72 points in total). For each stationary point, we performed a calculation of harmonic vibrational frequencies at the same level of theory, to properly characterize the minima and transition states (TSs), and to determine the ZPE, which is accounted for in the $E_{\rm iso}$ value. This also enabled us to ensure that the TSs indeed include only one imaginary frequency related with the expected reaction (interconversion) path.
%

%

\section{\rm \bf \large $\blacksquare$  STEREOISOMERS DETECTED IN THE ISM}
\label{sec-observations}

\begin{figure}
\begin{center}
\includegraphics[width=7.1cm,angle=-90]{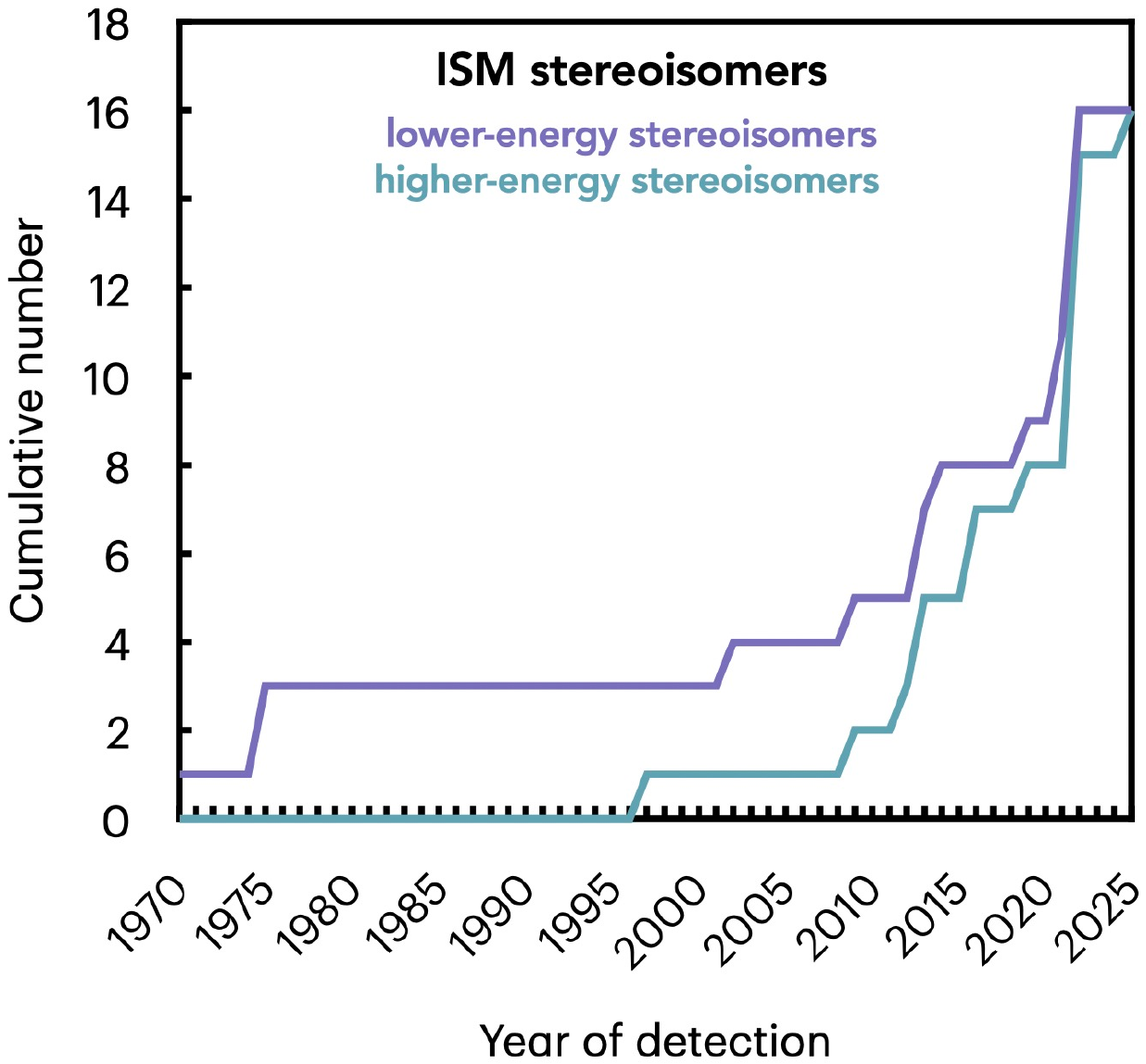}
\vspace{-5mm}
\end{center}
\caption{Cumulative number of stereoisomeric pairs detected as a function of time, considering the 16 interstellar stereoisomeric pairs discussed here. The lower(higher)-energy stereoisomers are indicated in purple(green), respectively.}
\label{fig-timeline}
\end{figure}

The rapid advances in sensitivity, bandwidth, and spectral resolution of modern astronomical facilities (e.g. the Yebes 40m and the Green Bank 100m single-dish telescopes, or the interferometer Atacama Large Millimeter/submillimeter Array, ALMA) have allowed the detection of higher-energy stereoisomers of previously known interstellar species.
%
%
Figure \ref{fig-timeline} shows the cumulative number of lower- and higher-energy stereoisomers belonging to the
stereoisomeric pairs identified so far in the ISM. 
In most cases, with only one exception\citep{Belloche2009,Belloche2014} (which will be discussed below), the lower-energy stereoisomers were detected first.
The first pair was completed in 1997, when the higher-energy isomer of ethanol (\ce{CH3CH2OH}) was detected towards the massive-star-forming region Orion KL\cite{Pearson1997}. More than 15 years later, the second and third higher-energy stereoisomers of previously known interstellar species were identified: ethyl formate (\ce{CH3CH2OCHO}) towards Orion KL\cite{Tercero2013}, and ethanimine (\ce{CH3CHNH}) towards the Sgr B2(N) hot core\cite{Loomis2013}. 
Since then, there has been a steady increase in the detection of higher-energy stereoisomers, with an accelerated 
rate in the last few years.

It is remarkable that half of them have been identified (through the first detection of the higher-energy isomer or the simultaneous detection of both isomers) since 2022 (Figure \ref{fig-timeline}).
As a fact, the gap between the detection of the lower and the higher-energy isomers has been narrowed dramatically. While the gap was 45 years for formic acid (HCOOH) \cite{Cuadrado2016,Zuckerman1971}, and similarly 37 and 22 years for methyl formate (\ce{CH3OCHO}) \cite{Brown1975,Neill2012} and ethanol (\ce{CH3CH2OH}) \cite{Zuckerman1971,Pearson1997}, in recent years it is becoming common that both isomers are discovered very closely in time, e.g., only separated by 1 year for thioformic acid (HCOSH)\cite{Rodriguez-Almeida2021,GarciadelaConcepcion2022} and vinyl alcohol (\ce{H2CCHOH})\cite{Agundez2021,Jimenez-Serra2022}, or even simultaneously, such as for ethanimine (\ce{CH3CHNH})\cite{Loomis2013}, $n$-propanol ($n$-{\ce{CH3CH2CH2OH}})\cite{Jimenez-Serra2022}, $i$-propanol  ($i$-\ce{(CH3)2CHOH})\cite{Belloche2022}, allyl cyanide (\ce{CH2CHCH2CN}) or crotononitrile  (\ce{CH3CHCHCN})\cite{Cernicharo2022}. 
Only in one case, $n$-propyl cyanide (\ce{CH3CH2CH2CN}), the high-energy isomer was reported 5 years earlier than its lower-energy counterpart\cite{Belloche2019,Belloche2014}, although this might be due to an incorrect energy ordering of the two stereoisomers assumed when performing the interstellar search (see further discussion in the next section).

\begin{figure*}
\begin{center}
\includegraphics[width=16.75cm]{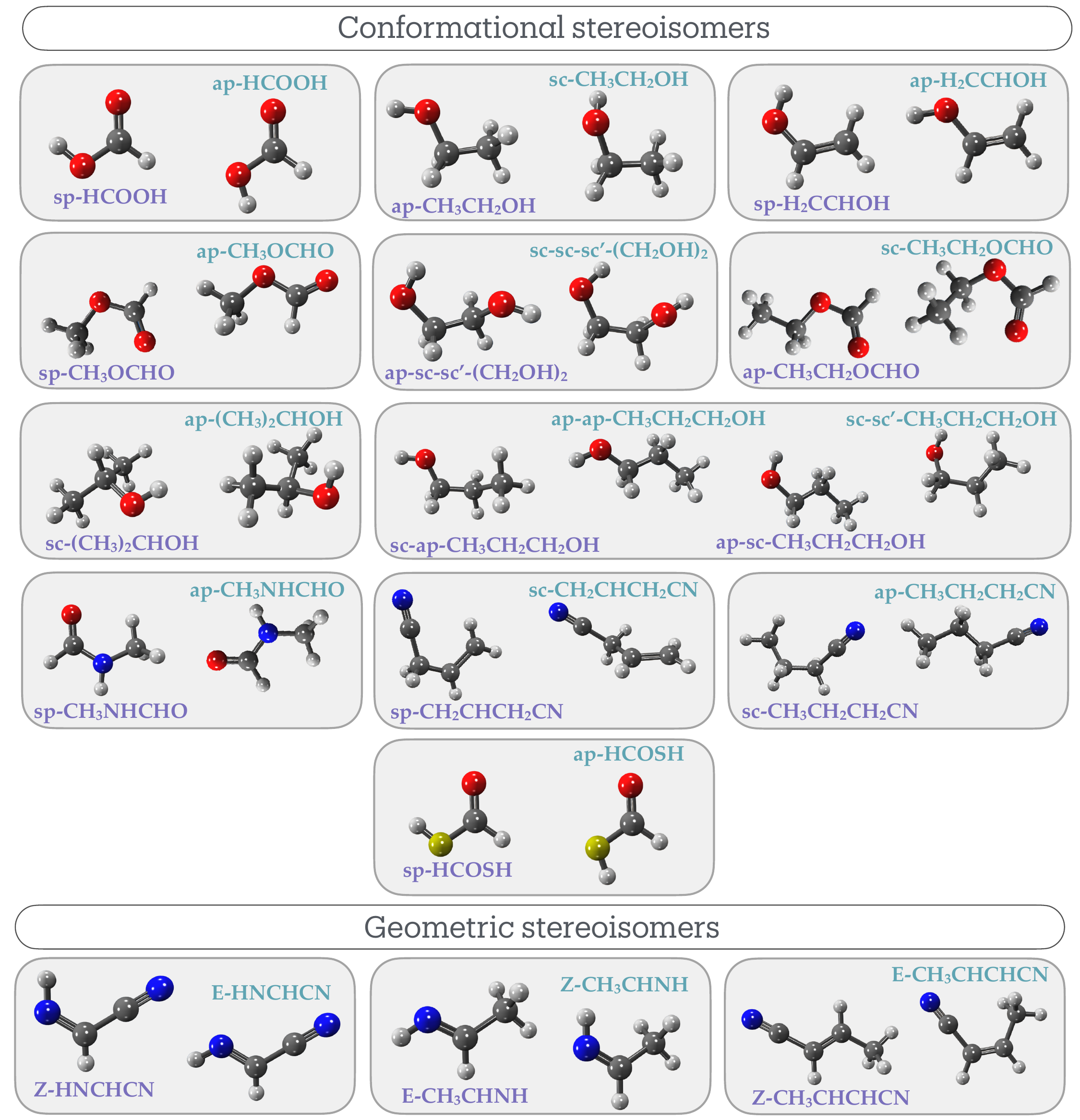}
\end{center}
\caption{Stereoisomers (separated into conformational and geometric, see Fig. \ref{fig-diagram}) detected in the interstellar medium. The purple(green) labels indicate lower(higher)-energy stereoisomers. 
The labels indicate the recommended nomenclature for the different stereoisomers, as discussed in the text. To check the correspondence with previous terminology, we refer to Table \ref{tab:stereoisomer_pairs}.
Gray, red, blue, yellow and white colors corresponds to atoms of carbon, oxygen, nitrogen, sulfur and hydrogen, respectively. The 3D representations of all molecules, optimized at the B3LYP-GD3/aug-cc-pVTZ level of theory using the Gaussian 16 program package,\cite{Frisch2016} were visualized with IQmol (\url{https://www.iqmol.org/}).}
\label{fig-3D}
\end{figure*}


\onecolumn

\setlength{\tabcolsep}{0.3pt} 
\renewcommand{\arraystretch}{0.66} 
{\footnotesize
\begin{landscape}
\begin{center}
\begin{longtable}{c c c c c c c c c c}
 \caption{Stereoisomeric pairs detected in the ISM.} \\
\hline \hline
Formula & Stereoisomer & $\Delta E$$^{(b)}$ & $E_{\rm iso}$$^{(c)}$ & $\mu$ & First ISM & OSR$^{(d)}$ (source) & \multicolumn{3}{c}{Refs.$^{(e)}$}  \\  \cline{3-4} \cline{8-10}
(name) & pair$^{(a)}$ &  \multicolumn{2}{c}{K [kcal mol$^{-1}$]}  & (Debye) & detection &  & $\Delta E$, $E_{\rm iso}$ & $\mu$ & Obs. \\
& (low/high)  &  &  &  (low/high) & (low/high) & (high/low) &  & (low/high) & (low/high) \\
\hline
\endfirsthead
\caption[]{Stereoisomeric pairs detected in the ISM (cont.)} \\
\hline \hline
Molecule & Stereoisomer & $\Delta E^{(b)}$ & $E_{\rm iso}$$^{(c)}$ & $\mu$ & First ISM & OSR$^{(d)}$ (source) & \multicolumn{3}{c}{Refs.$^{(e)}$}  \\  \cline{3-4} \cline{8-10}
& pair$^{(a)}$ & \multicolumn{2}{c}{K [kcal mol$^{-1}$]} &(Debye) & detection &  & $\Delta E$, $E_{\rm iso}$ & $\mu$ & Obs. \\
& (low/high)  &  & &  (low/high) & (low/high) & (high/low) &  & (low/high) & (low/high) \\
\hline
\endhead
\multicolumn{9}{c}{Conformational stereoisomers  (conformers) }  \\  
\hline
\ce{HCOOH} & trans/cis & 2033 [4.0] & 3687 [7.3] & 1.44/3.79 & \citep{Zuckerman1971}/$-$ & $-$ (Sgr B2(N) cor.) & \citep{GarciadelaConcepcion2022}, \citep{GarciadelaConcepcion2022} & \citep{Kuze1982}/\citep{Hocking1976} & $-$ \\
(formic acid) & (sp/ap) &  & &  &  $-$/\citep{Cuadrado2016} & 0.35(13) (Orion Bar) &  &  & \citep{Cuadrado2016} \\ 
 &  & & &  &  & 0.065(24) (B5) &  &  & \citep{Taquet2017} \\ 
 &  & & &  &  & 0.06(4) (L483) &  &  & \citep{Agundez2019} \\ 
 &  & & &  &  & 0.009(2) (G+0.693) &  &  & \citep{Sanz-Novo2023} \\ 
 &  & &  &  &  & 0.057(6) (TMC-1) &  &  & \citep{Molpeceres2025} \\ 
\hline
\ce{CH3CH2OH} & anti/gauche  & 57 [0.1] & 469 [0.93]  & 1.44/1.68 & \citep{Zuckerman1975}/$-$ & $-$ (Sgr B2) & \citep{Pearson2008,nandi2022}  & \citep{Takano1968}/\citep{Kakar1980}  & $-$ \\
(ethanol) & (ap/sc)  & &  &  & $-$/\citep{Pearson1997} & 4(3) (Orion-KL)  &  &  & \citep{Ohishi1995,Pearson1997} \\ 
\hline
\ce{H2CCHOH} & syn/anti & 541(75) [1.08(0.15)] &  2516 [5.0] & 1.02/1.79 & \citep{Agundez2021}/$-$ & $-$ (TMC-1) & \citep{Rodler1985,Rosch23} & \citep{Saito1976}/\citep{Rodler1985} & $-$ \\
(vinyl alcohol) & (sp/ap) & & &  & $-$/\citep{Jimenez-Serra2022} & 0.12(4) (G+0.693) &  &  & \citep{Jimenez-Serra2022} \\ 
\hline
\ce{CH3OCHO} & cis/trans & 2667 [5.3] & 6945 [13.8] & 1.77/4.89 & \citep{Brown1975,Neill2012} & 0.08(3) (Sgr B2(N) env.) & \citep{Senent2005}, \citep{Senent2005} & \citep{Curl1959}/\citep{Neill2012} & \citep{Faure2014}/ \citep{Neill2012}  \\ 
(methyl formate) & (sp/ap) & &  &  &  & 0.014(3) (G+0.693) &  &  & \citep{Sanz-Novo2025} \\ 
&  &  &  & & & 0.029(4) (L1157-B1) &  &  & \citep{Sanz-Novo2025} \\ 
\hline
\ce{(CH2OH)2} & aGg$'$/gGg$'$ & 300(48) [0.6(1)] & 740 [1.5] & 2.32/2.36 & \citep{Hollis2002}/$-$ & $-$ (Sgr B2(N) cor.) & \citep{MullerChristen2004}, \citep{Christen2001} & \citep{MullerChristen2004} & $-$ \\
(ethylene glycol) & (ap-sc-sc$^{\prime}$/sc-sc-sc$^{\prime}$) & & &  & $-$/\citep{Jorgensen2016} & 0.91(45) (IRAS 16293B) &  &  & \citep{Jorgensen2016} \\ 
 &  &  & &  &  & 0.40(16) (Orion-KL) &  &  & \citep{Favre2017} \\ 
 &  &  & &  &  & 0.60(14) (G31) &  &  & \citep{Mininni2023} \\ 
 &  &  & &  &  & 0.62(17)$^{(f)}$ (CoCCoA) &  &  & \citep{Chen2023}  \\ 
\hline
\ce{CH3CH2OCHO} & anti/gauche & 94(30) [0.19(6)] & 348 [0.69]$^{*}$ & 1.80/1.97 & \citep{Belloche2009}/$-$ & $-$ (Sgr B2 (N)) & \citep{Riveros1967},tw & \citep{Riveros1967} & $-$ \\ 
(ethyl formate) &  (ap/sc) & &  &  & $-$/\citep{Tercero2013}  &  1.0(3) (Orion-KL) & &  & \citep{Tercero2013} \\ 
 &  & & &  &  & 1.0(2) (W51 e2) &  &  & \citep{Rivilla2017} \\ 
\hline
\ce{(CH3)2CHOH} & gauche/anti & 120(6) [0.238(12)] & 468 [0.93] & 1.56/1.58 & \citep{Belloche2022}, \citep{Belloche2022} & 0.29(4)$^{(g)}$ (Sgr B2(N) cor.) & \citep{Maeda2006}, \citep{Salah2024} & \citep{Hirota1979} & \citep{Belloche2022} \\
($i$-propanol) & (sc/ap) & &  &  &  &  &  &  & \\
\hline
\ce{CH3CH2CH2OH} & Ga/Aa (sc–ap/ap-ap) & 40.3 [0.08] &  1566 [3.1]$^{*}$ & 1.48/1.46 & \citep{Jimenez-Serra2022},\citep{Jimenez-Serra2022} & 0.62(7) (G+0.693) &  \citep{KahnBruice2005},tw & \citep{Kisiel2010} & \citep{Jimenez-Serra2022} \\ 
($n$-propanol) & Ag/Gg$'$ (ap–sc/sc–sc$'$)  & 10 [0.02] &   1733 [3.4]$^{*}$ & 1.68/1.65 & \citep{Belloche2022},\citep{Belloche2022} & 0.96(14)$^{(g)}$ (Sgr B2(N) cor.) &  &  & \citep{Belloche2022} \\ 
\hline
\ce{CH3NHCHO} & trans/cis & 705 [1.4] & 9848 [19.6]$^{*}$ & 3.90/4.47 & \citep{Belloche2017},\citep{Belloche2019}/$-$ & $-$ (Sgr B2(N) cor.) &  \citep{Zeng2025},tw & \citep{Lattelais2010}/\citep{Kawashima2010} & $-$ \\ 
(N-methyl formamide) &  (sp/ap) &  & &  & $-$/\citep{Zeng2025} & 0.34(7) (G+0.693) &  &  & \citep{Zeng2023}/\citep{Zeng2025} \\ 
\hline
\ce{CH2CHCH2CN} & cis/gauche  & 96 [0.19] & 1261 [2.5]$^{*}$ & 3.91/3.98 & \citep{Cernicharo2022}/\citep{Cernicharo2022} & 1.14(16) (TMC-1) & \citep{Cernicharo2022},tw & \citep{Sastry1968}  & \citep{Cernicharo2022} \\ 
(allyl cyanide) &  (sp/sc) &  &  & &  &  & &  &  \\
\hline
\ce{CH3CH2CH2CN} & gauche/anti  &  58(5) [0.11(1)] & 1749 [3.5] &  3.93/4.12 &  \citep{Belloche2014}/\citep{Belloche2009} &  0.39(6)$^{(g)}$ (Sgr B2(N) cor.) &  \citep{Durig2001},\citep{Kerkeni2019} &  \citep{Durig2001},\citep{Wlodarczak1988}/\citep{Wlodarczak1988},\citep{Muller2011} &  \citep{Belloche2014},\citep{Muller2016}  \\ 
 ($n$-propyl cyanide) & (sc/ap)  &  &  &  &  &  0.36(5)$^{(g)}$ (Orion-KL) &  &  &  \citep{Pagani2017} \\
\hline
\ce{HCOSH} & trans/cis & 342 [0.68] & 3996 [7.9] &1.54/2.87 & \citep{Rodriguez-Almeida2021}/$-$ & $\leq$0.2 (G+0.693) & \citep{GarciadelaConcepcion2022},\citep{GarciadelaConcepcion2022} & \citep{HockingWinnewisser1976} & \citep{Rodriguez-Almeida2021} \\ 
(thioformic acid) & (sp/ap) & &  &  & $-$/\citep{GarciadelaConcepcion2022} & 0.27(11) (G31) &  &  & \citep{GarciadelaConcepcion2022} \\ 
\hline
\multicolumn{9}{c}{Configurational stereoisomers (geometric isomers)}   \\  
\hline
\ce{HNCHCN} & Z/E & 309(72) [0.61(14)] & 13335 [26.5]  & 1.46/4.56 & \citep{Rivilla2019}/$-$ & 0.22(1) (G+0.693) & \citep{Takano1990},\citep{Puzzarini2015} & \citep{Puzzarini2015} & \citep{SanAndres2024} \\ 
(C-cyanomethanimine) &  & &  &  & $-$/\citep{Zaleski2013} & $-$ (Sgr B2(N) cor.) & & & $-$ \\
\hline
\ce{CH3CHNH} & E/Z & 327 [0.65] & 13753[27.3] & 2.04/2.42 & \citep{Loomis2013}/\citep{Loomis2013} & 0.33(7) (Sgr B2(N) cor.) & \citep{GarciadelaConcepcion2021},\citep{GarciadelaConcepcion2021}  & \citep{Lovas1980}/\citep{Brown1980}  & \citep{Loomis2013} \\ 
(ethanimine) &  & &  &  &  & 0.085(15) (G+0.693) &  &  & \citep{GarciadelaConcepcion2021} \\ 
\hline
\ce{CH3CHCHCN} & trans/cis & 1143 [2.3] & 25815 [51.3] &4.75/4.08 & \citep{Cernicharo2022}/\citep{Cernicharo2022} & 2.6(5) (TMC-1) & \citep{Cernicharo2022},\citep{Butler1963} & \citep{SuzukiKozima1970}/\citep{Beaudet1963} & \citep{Cernicharo2022} \\ 
(crotononitrile) & (Z/E) & &  & &  &  & &  &  \\
\hline
\label{tab:stereoisomer_pairs}
\end{longtable}
{(a) We include the nomenclature usually employed in the literature, while the formally correct one (recommended by IUPAC) is given in parentheses (see text);
(b) $\Delta E$ values for which an uncertainty is provided correspond to spectroscopic measurements, whereas the remaining come from \textit{ab initio} calculations. As a distinctive case, the $\Delta E$ for \ce{CH3CH2OH} was determined from the analysis on the torsional potential of this species \citep{Pearson2008}. Numbers in parentheses indicate the uncertainty on the last digits;
(c) Values marked with $^{*}$ were derived from the relaxed potential energy surface calculated in this work by varying the corresponding dihedral angle (at the B3LYP-GD3/aug-cc-pVTZ level of theory, including ZPE corrections);
(d) OSR = Observed Stereoisomeric Ratio (higher/lower).  Numbers in parentheses indicate the uncertainty on the last digits; 
(e) tw: this work;
(f) Average value calculated for a sample of 11 hot cores within the CoCCoA survey\cite{Chen2023};
(g) OSR calculated directly by assuming thermodynamic equilibrium between both stereoisomers. 
For all these particular cases, an uncertainty of 15\% of the OSR value has been considered.
}
\end{center}
\end{landscape}
}

\twocolumn

The total number of stereoisomeric pairs detected so far is 16, which are listed in Table \ref{tab:stereoisomer_pairs}. 
Among them, 13 are conformational (including two pairs of $n$-propanol, \ce{CH3CH2CH2OH}), and 3 are geometric.
The 3-dimensional representations of these stereoisomers computed in this work (see previous section) are shown in Fig. \ref{fig-3D}. 
%
%
%
They are molecules with 5 to 12 atoms.
All of them contain carbon and hydrogen, 10 are oxygen-bearing, 6 are nitrogen-bearing and 1 is sulfur-bearing. 
Table \ref{tab:stereoisomer_pairs} also summarizes their total dipole moments ($\mu$), and the year of the first discovery of each isomer in the ISM. 

For further discussion in the next section, Table \ref{tab:stereoisomer_pairs} also presents the relative isomeric energy difference ($\Delta E$), and the isomerisation energy barriers ($E_{\rm iso}$). 
For the isomers of \ce{H2CCHOH}, \ce{(CH2OH)2}, \ce{CH3CH2OCHO}, \ce{(CH3)2CHOH}, \ce{CH3CH2CH2CN} and \ce{HNCHCN}, the reported $\Delta E$ are experimental values based on spectroscopic measurements, and their statistical uncertainties (associated to the relative intensity measurements) are given in Table \ref{tab:stereoisomer_pairs}. For the remaining species, $\Delta E$ has been determined through \textit{ab initio} calculations reported in the literature.
%
%
In the case of $E_{\rm iso}$, with the exception of crotonitrile, the values come in all cases from theoretical calculations from previous works, or newly performed in this work for the stereosiomers of \ce{CH3CH2OCHO}, \ce{CH3CH2CH2OH}, \ce{CH3NHCHO}, and \ce{CH2CHCH2CN}.

%

We note that the theoretical values of $\Delta E$ and $E_{\rm iso}$ 
should be interpreted within the typical uncertainty of the underlying quantum-chemical calculations. While the accuracy of DFT methods using presently-available functionals are limited to 2-3 kcal mol$^{-1}$ for many molecules,\cite{Bogojeski2020} $ab$ $initio$ methods such as coupled-cluster theory can routinely reach the so-called chemical accuracy of $\lesssim$1 kcal mol$^{-1}$ (i.e., $\lesssim$500 K).\cite{Raghavachari23} 
%

The three isomeric pairs with higher $\Delta E$ are methyl formate (2667\K), formic acid (2033\K) and  crotononitrile (1143\K), which clearly confirms that the detection of higher-energy stereoisomers is possible in the ISM, despite the previous general thought based on their thermodynamic instability. Regarding the values of $E_{\text{iso}}$, in all cases these are above 300\K (and even $>$1000\K for most of them), which in most cases prevents the isomerisation in gas phase at the temperatures of the ISM.

The interstellar stereoisomers have been detected towards different types of astronomical objects, listed in Table \ref{tab:OSR_sources_list}, which include dark molecular clouds, a photodissociation region, hot environments surrounding massive and low-mass protostars (hot cores and corinos, respectively), and shocked-dominated regions. 
These different sources span a wide range of gas kinetic temperatures ($T_\text{kin}$), from $\sim$7.5\K to $\sim$300\K (see references in Table \ref{tab:OSR_sources_list}).
This, together with the values of $\Delta E$, will allow us to study if the behaviour of the stereoisomeric ratio follows the expectations of thermodynamics (according to the MEP), or alternatively there are significant deviations (see next section). 

\begin{figure}
\begin{center}
\includegraphics[width=8cm,angle=0]{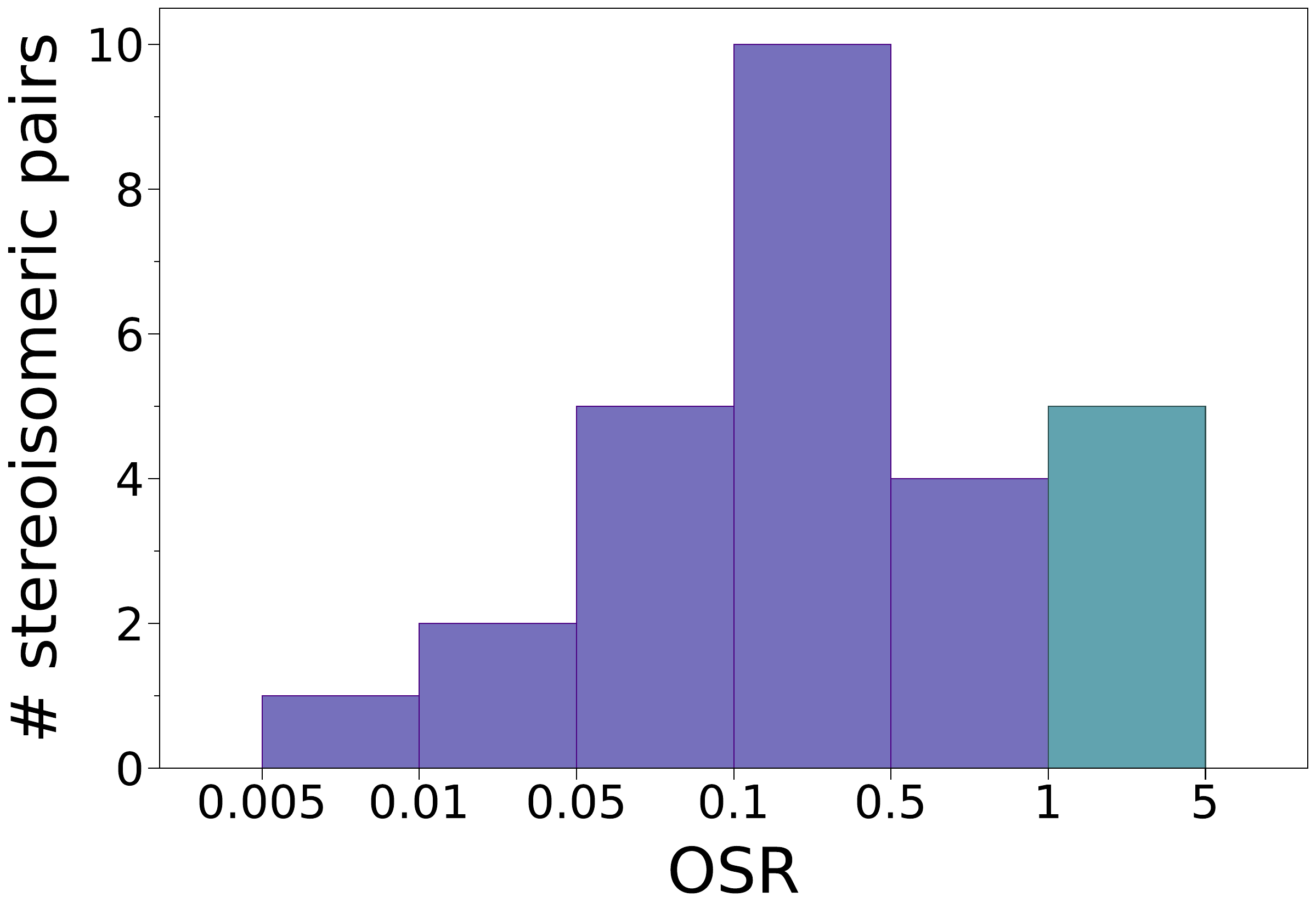}
\caption{Histogram of the values of the Observed Stereoisomeric Ratio (OSR) towards different sources of the ISM. Purple is used for values $<$ 1, and green for $\geq$1.}
\label{fig:OSR_histo}
\end{center}
\end{figure}

We defined the observed stereoisomeric ratio (hereafter OSR) as the column density ratio of the higher-energy isomer divided by that of the lower-energy isomer, and we reported the values derived towards the different astronomical sources in Table \ref{tab:stereoisomer_pairs}. 
The OSRs were computed using the column densities of the stereoisomers published in the references indicated in Table \ref{tab:stereoisomer_pairs}. For each isomeric pair, the observations used for the detection of each pair member (which are single-dish for some cases, and interferometric in others) are the same or at least consistent in terms of observational beams, and thus provide reliable ratios. The derived values of the OSRs are based on the reasonable assumption that both stereoisomers are co-spatial.
The distribution of the OSRs is shown in Figure \ref{fig:OSR_histo}.
The values 
span from 0.009 to 4. Although most of them are $<$ 1, there are a few exceptions in which the higher-energy isomer is equally abundant, such as ethyl formate, or even and more abundant, such as ethanol, allyl cyanide and crotononitrile (4$\pm$3 and 1.14$\pm$0.16, 2.6$\pm$0.5, respectively).

\setlength{\tabcolsep}{2.4pt} 
\renewcommand{\arraystretch}{1} 

\begin{table*}[]
    \centering
    \begin{tabular}{c c c c c c}
    \hline\hline
    Source & RA & DEC & Type & $T_\text{kin}$$^{(b)}$ &  Refs. \\
    & (ICRS) & (ICRS) &  & (K) & \\
    \hline
    B5 & \ra{03}{47}{32.10} & \dec{32}{56}{43.00} & Dark cloud & 7.5 & \citep{Taquet2017} \\
    TMC-1 & \ra{04}{41}{41.90} & \dec{25}{41}{27.10} & Dark cloud & 9 & \citep{Agundez2023} \\ 
    L483 & \ra{18}{17}{29.80} & \dec{-04}{39}{38.00} & Dark cloud$^{(a)}$ & 9.5 & \citep{Anglada1997},   \citep{Tafalla2000} \\ 
    Orion Bar & \ra{05}{35}{20.10} & \dec{-05}{25}{07.00} & Photodissociation region & 150  & \citep{BatrlaWilson2003} \\
    W51 e2 & \ra{19}{23}{43.90} & \dec{14}{30}{34.80} & Hot core & 150 & \citep{Remijan2004} \\
    Orion KL & \ra{05}{35}{14.50} & \dec{-05}{22}{30.00} & Hot core & 180 & \citep{Tang2018} \\
    G31.41+0.31 & \ra{18}{47}{34.27} & \dec{-01}{12}{43.20} & Hot core & 200 & \citep{Mininni2020}, \citep{Beltran2018} \\
    Sgr B2(N) cores & \ra{17}{47}{19.87} & \dec{-28}{22}{15.78} & Hot core & 225 & \citep{Moller2023} \\
    Sgr B2(N) envelope & \ra{17}{47}{19.80} & \dec{-28}{22}{17.00} & Hot core envelope & 30 & \citep{Faure2014} \\
    IRAS 16293-2422 B & \ra{16}{32}{22.58} & \dec{-24}{28}{32.80} & Hot corino & 300$^{(c)}$ &  \citep{Jorgensen2016} \\
    L1157-B1 & \ra{20}{39}{10.20} & \dec{68}{01}{10.50} & Protostellar shock & 70$^{(d)}$ & \citep{Lefloch2012},\citep{Gomez-Ruiz2015},\citep{Lefloch2016} \\
    G+0.693-0.027 & \ra{17}{47}{21.92} & \dec{-28}{21}{27.20} & Shocked molecular cloud & 140 & \citep{Zeng2020}, \citep{Colzi2024} \\
    \hline
    \end{tabular}
    \caption{List of sources where stereoisomers have been detected in the ISM.}
    \label{tab:OSR_sources_list}
    {$^{(a)}$ L483 is dark cloud with an embedded Class 0 protostar.
    $^{(b)}$ For most sources, the $T_\text{kin}$ shown here is the average of the values derived from the analyses performed on several tracers, such as \ce{CO}, \ce{NH3}, \ce{H2CO}, \ce{CH3CN} or \ce{CH3CCH} among others.
    $^{(c)}$ Excitation temperature ($T_\text{ex}$) derived for \ce{(CH2OH)2} stereoisomers (the only pair among those shown in this work detected towards this source), assuming the gas is thermalised (i.e., $T_\text{kin} = T_\text{ex}$). 
    $^{(d)}$ Average value derived for the $g_2$ velocity component among the three identified in the source, which is the one dominating the emission within the cavity of the bow shock (\citealt{Lefloch2012}).} 
\end{table*}

We provide below a description of the different interstellar stereoisomeric pairs:

{\it Conformational stereoisomers (conformers)}

{\bf $\bullet$ Formic acid (HCOOH)}: this species presents two conformational isomers, which formally should be named $sp$ and $ap$ (defined with respect to the priority bonds, i.e., the O=C-O-H dihedral angle), although they are commonly named $trans$ and $cis$ in the literature. The $trans$ ($sp$) conformer is strongly stabilized relative to the $cis$ ($ap$) form by an intramolecular O–H···O hydrogen-bonding interaction. In addition, this interaction contributes significantly to the height of the torsional barrier governing the interconversion between the two conformers, which involves rotation of the OH group around the O–C single bond, i.e. variation of the H–O–C–O dihedral angle (Figure \ref{fig-3D}). As the OH group rotates, the favourable intramolecular hydrogen-bonding interaction is progressively weakened, and substantial electronic rearrangement occurs within the carboxyl group. Consequently, both the energy difference and the isomerisation barriers are calculated to exceed 2000\K.\cite{GarciadelaConcepcion2022}

{\bf $\bullet$ Ethanol (\ce{CH3CH2OH})}: it exhibits two low-energy conformers, commonly referred to as $anti$ and $gauche$ (formally $ap$ and $sc$), which differ in the relative orientation of the hydroxyl group with respect to the C–C bond (Figure \ref{fig-3D}). The interconversion between both conformers occurs through rotation around the C–C single bond (i.e., varying the H-O-C-C dihedral angle), leading to a modest torsional barrier. As a consequence, the computed energy difference between the two forms is very small ($\Delta$$E$ = 57\K)\cite{Pearson2008}, 
and the isomerisation barrier is also low (469\K)\cite{nandi2022}.

{\bf $\bullet$ Vinyl alcohol (\ce{H2CCHOH})}: this molecule exists as two conformational isomers, $syn$ and $anti$ ($sp$ and $ap$), depending on the orientation of the OH with respect to the C=C double bond. As shown in Figure \ref{fig-3D}, the OH group is oriented towards the $\pi$ bond in the $syn$ ($sp$)  conformer,\footnote{A $\pi$ bond is a type of covalent bond formed by the sideways overlap of parallel p- (or d-) orbitals on adjacent atoms, creating an electron density above and below the plane of the nuclei. This bond restricts rotation around the bond axis, making double bonds much less flexible than single ($\sigma$) bonds.} whereas in the $anti$ ($ap$) form it points away from it. The interconversion between both conformers requires rotation of the OH group about the C–O bond (i.e., varying the H-O-C-C dihedral angle). The energy difference between the two conformers is measured to be relatively low ($\Delta E = 541\pm75$\K)\cite{Rodler1985}, while the torsional barrier reaches $E_{\rm iso} = 2516$\K\citep{Rosch23}, being significantly larger than that of ethanol. 

{\bf $\bullet$ Methyl formate (\ce{CH3OCHO}):} this species exhibits two planar conformers, commonly labeled as $cis$ and $trans$ 
(formally $sp$ and $ap$), which differ in the relative orientation of the methoxy and carbonyl groups (Figure \ref{fig-3D}). The interconversion between them involves rotation around the C–O single bond, coupled with significant torsional rearrangement of the molecular framework. This process is associated with a large energy difference, calculated to be $\Delta E = 2667$\K, and a very high isomerisation barrier of $E_{\rm iso} = 6945$\K, owing to a significant electrostatic repulsion related to the rotation of the methyl group.\citep{Senent2005} Such a high barrier effectively prevents reaching thermal equilibrium between the conformers under typical interstellar conditions, making methyl formate the most extreme case among detected conformational isomeric pairs.

{\bf $\bullet$ Ethylene glycol (\ce{(CH2OH)2}):} it displays several conformational minima, with the two lowest-energy structures, $aGg'$ and $gGg'$, characterized by the presence of an intramolecular hydrogen bond and a $gauche$ arrangement of the OCCO backbone. Here, the capital letters $G$ and $A$ refer to the rotation of the heavy nuclei plane C–C–C–C along the OCCO backbone, with $A$ corresponding to $anti$ and $G$ to $gauche$, while the small letters $g$ and $a$ describe the rotation of the hydroxy group (-OH) dihedral angle. 
According to the recommended nomenclature, these conformer should be named $ap$-$sc$-$sc$$'$ and $sc$-$sc$-$sc$$'$, respectively.
The prime ($^{\prime}$) indicates a slightly different torsional variant. In both conformers, the dominant stabilization arises from the favorable intramolecular hydrogen bond between the two OH groups, leading to a folded geometry (see Figure \ref{fig-3D}). Their  measured energy difference is $\Delta E = 300\pm48$\K{\cite{MullerChristen2004},
while the isomerisation barrier is $E_{\rm iso} = 720$\K{\cite{Christen2001}. The interconversion between them primarily involves rotation around the central C–C single bond (i.e., varying the O-C-C-O dihedral angle), leading to changes in the relative orientation of the two OH groups while preserving the overall folded geometry of the molecule.

{\bf $\bullet$ Ethyl formate (\ce{CH3CH2OCHO}):} this molecule exhibits two low-energy conformers associated with different orientations of the ethyl group relative to the ester moiety. In the most stable conformer, the terminal CH$_3$ group is oriented $anti$ ($ap$) to the O–CH$_2$ bond, whereas in the higher-energy $gauche$ ($sc$) form it is rotated by approximately 60° (Figure \ref{fig-3D}). The experimentally-determined energy difference between the two conformers is $\Delta E = 94\pm30$\K\cite{Riveros1967} 
, and the interconversion barrier we derived is very low ($E_{\rm iso} = 342$\K).
This conformational change involves rotation around the O–C single bond of the O–CH$_2$–CH$_3$ fragment (i.e., varying the C-O–C–C dihedral angle).

{\bf $\bullet$ $i$-propanol or 2-propanol (\ce{(CH3})$_2$CHOH):} the two low-energy conformers, $gauche$ and $anti$ ($sc$ and $ap$), are very close in energy ($\Delta E = 120\pm6$\K, determined through spectroscopic measurements\cite{Maeda2006}).
As shown in Figure \ref{fig-3D}, they differ in the orientation of the OH relative to the central C–C bond. Interconversion occurs via rotation around this C–C single bond (i.e., varying the H–O–C–C dihedral angle), with a low torsional barrier of $E_{\rm iso} = 468$\K\cite{Salah2024}, allowing interconversion even at low temperatures. 


{\bf $\bullet$ $n$-propanol or 1-propanol (\ce{CH3CH2CH2OH})}: the two observed pairs of conformers, Ga/Aa and Ag/Gg$'$ (or $sc$-$ap$/$ap$-$ap$ and $ap$–$sc$/$sc$–$sc$$'$), exhibit low energy differences of $\Delta E = 40.3$\K and $\Delta E = 10$\K, respectively \cite{KahnBruice2005}. 
In this case, the capital letters $G$ ($gauche$) and $A$ ($anti$) refer to the rotation of the heavy nuclei plane C–C–C relative to the C–C–O plane, with $A$ corresponding to $\sim$180° and $G$ to $\sim$60°, while the small letters $g$ and $a$ describe the rotation of the hydroxy group (-OH) dihedral angle. Therefore, the conformers differ in the relative orientation of the OH and CH$_3$ groups along the carbon chain. Thus, conformational interconversion involves rotation around the central and terminal C–C single bonds (i.e., varying the O–C–C–C and H–C–C–C dihedral angles, respectively), with torsional barriers of $E_{\rm iso} = 1566$\K ($sc$-$ap$/$ap$-$ap$) and $E_{\rm iso} = 1733$\K ($ap$–$sc$/$sc$–$sc$$'$) as derived in this work. Figure \ref{fig-3D} illustrates the different conformations, highlighting the more folded (compact) versus extended geometries resulting from the distinct hydroxyl and methyl group orientations.

{\bf $\bullet$ N-methyl formamide (\ce{CH3NHCHO})}: this molecule shows $trans$-$cis$ isomerism (formally $sp$-$ap$), as the molecule does not exhibit a double bond or rigid framework about the amide C–N bond (OC–NH; see Figure \ref{fig-3D}). Due to the partial double-bond character of the amide linkage, the rotation around this bond is strongly hindered. The so-called $trans$ conformer is more stable, with a calculated energy difference of $\Delta E = 705$\K\cite{Zeng2025} relative to the $cis$ conformer, while the torsional barrier for interconversion (i.e., rotation around the C–N single bond) we derive is 
high, $E_{\rm iso} = 9848$\K, owing to steric hindrance between the carbonyl and the CH$_3$ groups.

{\bf $\bullet$ Allyl cyanide (\ce{CH2CHCH2CN})}: this species, also known as 3-butenenitrile, 2-propenenitrile or 2-cyanopropene, exhibits conformational isomerism due to rotation around the single bond connecting the allyl and nitrile groups. The two low-energy conformers, commonly labeled as $cis$ (formally it should be $sp$) and $gauche$ ($sc$), differ in the orientation of the –\ce{CH2CN} group relative to the vinyl moiety (Figure \ref{fig-3D}), and they are separated by $\Delta E = 96$ K (from \textit{ab initio} calculations\cite{Cernicharo2022}).
The torsional barrier computed in this work is $E_{\rm iso} = 1261$\K, and the interconversion occurs through rotation about the C–C–C–C dihedral angle, affecting the spatial arrangement of the allyl and nitrile groups while preserving the characteristic linearity of the \ce{CH2CN} fragment.

{\bf $\bullet$ {\it n}-propyl cyanide ({\it n-}butanenitrile or {\it n-}butyronitrile; \ce{CH3CH2CH2CN}):} this molecule exhibits two low-energy conformers, usually cited as $gauche$ and $anti$, which formally should be named $sc$ and $ap$. They differ in the relative orientation of the carbon backbone. Initially, it was thought that the the $anti$ ($ap$) conformer was lower in energy than the $gauche$ ($sc$) conformer\cite{Wlodarczak1988}, but later works have confirmed that the relative energy ordering is reverse\citep{Durig2001,Kerkeni2019}, with the $gauche$ ($sc$) conformer being the global minimum in energy. 
The measured energy difference is $\Delta E = 58\pm5$\K \citep{Durig2001}, and the computed torsional barrier is $E_{\rm iso} = 1749$\K\cite{Kerkeni2019}.


{\bf $\bullet$ Thioformic acid (\ce{HCOSH})}: this species presents a situation analogous to formic acid, with two conformers commonly referred to as $trans$ and $cis$ ($sp$ and $ap$, as in the case of HCOOH), depending on the orientation of the SH group relative to the carbonyl oxygen (Figure \ref{fig-3D}). The $sp$ conformer is moderately stabilized by a weak intramolecular S–H···O hydrogen-bonding interaction, being  342\K below the $ap$ form according to theoretical calculations\cite{GarciadelaConcepcion2022}. 
Despite the relatively modest energy difference between the two minima, the isomerisation barrier is high ($E_{\rm iso} = 3996$\K)\cite{GarciadelaConcepcion2022}, exceeding that of formic acid. The interconversion proceeds via rotation around the O–C–S–H dihedral angle, with the torsional motion controlling the relative orientation of the SH group and the carbonyl moiety, and strongly perturbing the electronic structure of the molecule.


\vspace{5mm}

{\it Geometric stereoisomers}

{\bf $\bullet$ C-cyanomethanimine  (\ce{HNCHCN})}: this molecule exists as two geometric stereoisomers, $Z$ and $E$, defined by the orientation around the C=N double bond. According to the Cahn–Ingold–Prelog (CIP) priority rules, the $Z$ isomer has the highest-priority substituents on the same side of the double bond, while in the $E$ isomer they are on opposite sides (see Figure \ref{fig-3D}).  Rotation around the C=N bond is highly restricted under typical thermal conditions, effectively “freezing” the isomers at low temperatures. Interconversion requires breaking the $\pi$ component of the double bond, which results in a very high isomerisation barrier ($E_{\rm iso} = 13335$\K)\cite{Puzzarini2015}. Nevertheless, the energy difference between the two forms is relatively low ($\Delta E = 309\pm72$\K, determined through spectroscopic measurements\citep{Takano1990}).

{\bf $\bullet$ Ethanimine (\ce{CH3CHNH})}: Figure \ref{fig-3D} depicts the two plausible geometric isomers of ethanimine, $E$ and $Z$, which differ in the relative orientation of the substituents across the C=N double bond. As in C-cyanomethanimine, interconversion between the isomers requires breaking the $\pi$ bond, which requires substantial energy. Consequently, although the calculated energy gap between these isomers is very small ($\Delta E = 327$\K), the interconversion barrier is extremely high ($E_{\rm iso} = 13753$\K)\cite{GarciadelaConcepcion2021}. 

{\bf $\bullet$ Crotononitrile (\ce{CH3CHCHCN})}: it exhibits two geometric isomers, commonly labeled in this case as $trans$ and $cis$ (formally $Z$ and $E$), which describe the relative disposition of the substituent with respect to the C=C double bond (see Figure \ref{fig-3D}). As with other double-bond systems, rotation is highly restricted because it requires partial disruption of the $\pi$ bond, leading to a very high interconversion barrier ($E_{\rm iso} = 25815$\K, determined in a kinetic study by \cite{Butler1963} in which thermochemistry is also analyzed). The $Z$ isomer is considerably more stable than the $E$ form, with a calculated energy difference of $\Delta E = 1143$\K\cite{Cernicharo2022}, as the latter experiences greater steric hindrance due to the positioning of the CH$_3$ group.

\section{\rm \bf \large $\blacksquare$  DISCUSSION: THE ORIGIN(S) OF THE OSR}
\label{sec-discussion}

The relatively large sample of stereoisomers detected so far in the ISM allow us to study systematically their isomeric ratios, and to search for possible general trends that help to rationalize the stereoisomeric chemistry, and, as a consequence, the detectability of higher-energy stereoisomers. 
In particular, we are now able to test
whether the OSRs follow the expectations of thermodynamic stability, or if they show significant deviations. In this section we discuss both possible scenarios, based on the results from astronomical observations.
We review the different mechanisms that can contribute to shape the stereoisomeric ratios, such as direct thermal isomerisation (i.e., without involving any other additional species), which can be assisted by 3-dimensional quantum tunneling, induced by radiation or by energy input during desorption from grain mantles; and stereoselective formation/destruction chemical reactions. A summary of these mechanisms, along with the works in which have been proposed and discussed, is provided in Table \ref{tab:mechanisms}.




%

As shown in Figure \ref{fig:OSR_histo}, most of the
OSRs are $<$1 (i.e., the higher-energy stereoisomer is less abundant), which
might point the MEP to be a valid rule.
If the OSRs were exclusively governed by thermodynamics, their values would then be described by:

\begin{equation}
{\rm OSR} = \frac{N_{\rm higher}}{N_{\rm lower}} = g \times exp(-\Delta E/T_\text{kin}),
\label{eq-OSR}
\end{equation}

where $\Delta E$ is the energy difference between the isomers, $T_\text{kin}$ the kinetic temperature of the gas, and $g$ accounts for the possible degeneracy of the members of the stereoisomeric pair: $g$=1 if there is not degeneracy, $g$=(1/2) if the lower-energy isomer is double degenerated, and $g$=2 is the higher-energy isomer is doubly degenerated.
For the stereisomers discussed here, $g$=1 for all the cases, with the exception of \ce{CH3CH2OH}, \ce{CH3CH2OCHO}, \ce{CH2CHCH2CN}, \ce{CH3CH2CH2CN} and \ce{(CH3)2CHOH} in which one of the isomers in the pair is doubly degenerated (typically the $gauche$, i.e., the $sc$ conformer): the higher-energy for the first three (and then $g$=2), and the lower-energy for the last two ($g$=1/2).

\begin{figure*}
\begin{center}
\includegraphics[width=17.5cm,angle=0]{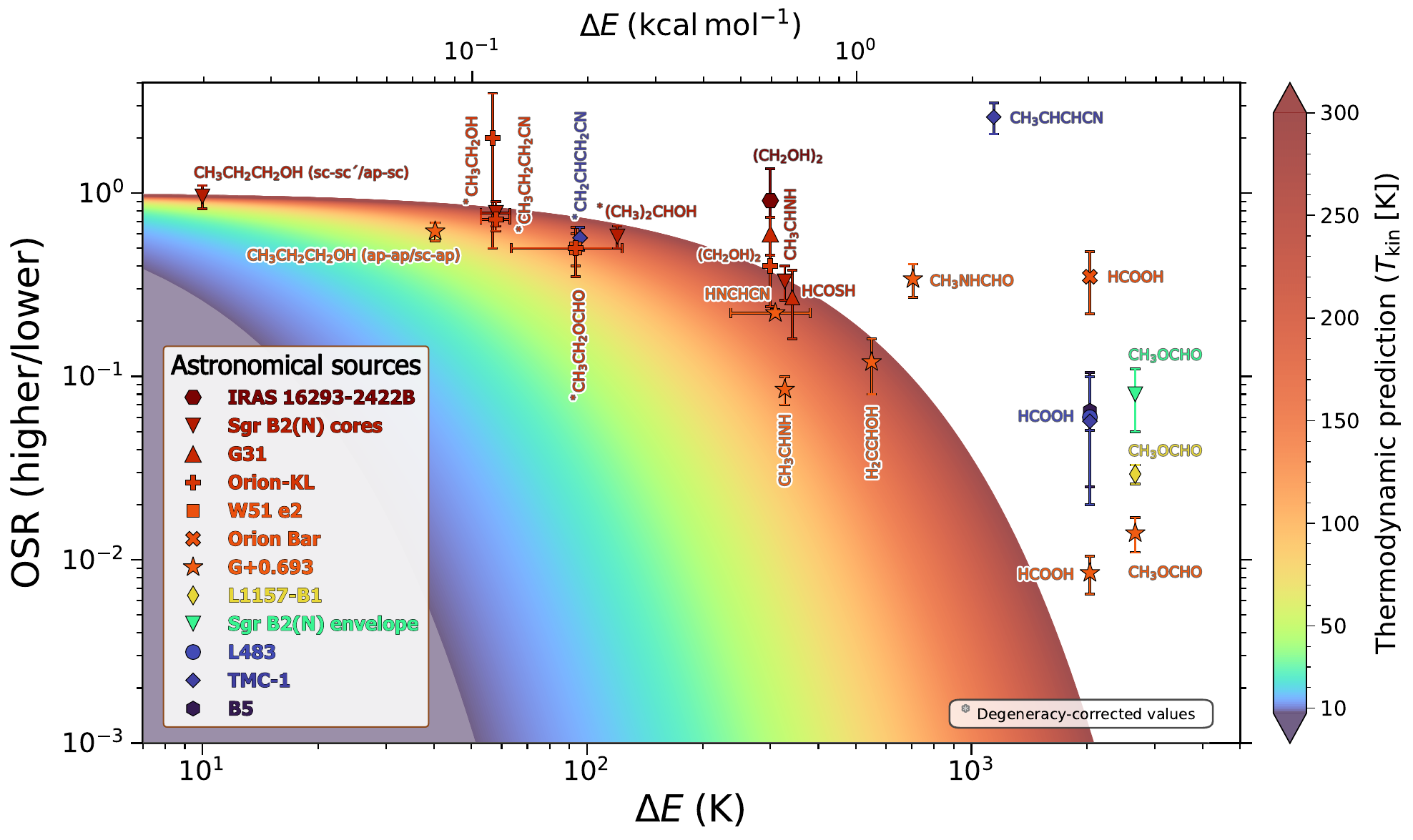}
\caption{Observed Stereoisomeric Ratio (OSR) - defined as the column density ratio between the higher and the lower-energy isomer - for the different stereoisomeric pairs gathered in Table~\ref{tab:stereoisomer_pairs} (points), overlaid with the thermodynamic equilibrium prediction ($g \times e^{-\Delta E/T_\text{kin}}$ at $g = 1$, colored area), as a function of the relative energy ($\Delta E = E_\text{higher}$ - $E_\text{lower}$) between isomers. The points corresponding to the detections are colored according to the particular source $T_\text{kin}$ (see Tables~\ref{tab:stereoisomer_pairs} and ~\ref{tab:OSR_sources_list}), following the color code of the vertical bar. 
For \ce{CH3CH2OH}, \ce{CH3CH2OCHO}, \ce{CH2CHCH2CN}, \ce{CH3CH2CH2CN} and \ce{(CH3)2CHOH}, the observational points were ``corrected'' (indicated with *) accounting for the double degeneracy of one of the isomers in the pair (the higher-energy for the first three, the lower-energy for the last two), multiplying the corresponding ratios shown in Table~\ref{tab:stereoisomer_pairs} by $1/g$=1/2, 1/2, 1/2, 2 and 2, respectively, to match the displayed thermodynamic prediction which omits this fact. 
}
\label{fig:OSR_deltaE}
\end{center}
\end{figure*}

\begin{table*}[htbp]
\centering
\caption{Summary of isomerisation and stereoselective mechanisms.}
\label{tab:mechanisms}
\renewcommand{\arraystretch}{1.3}
\begin{tabularx}{0.945\linewidth}{|l|c|c|c|}
\hline
\multicolumn{3}{|c}{Mechanism}  & \multicolumn{1}{| c |}{Refs.} \\ \cline{2-4}
\hline
\multirow{2}{*}{\bf Stereoselective kinetics on grains}
& \multicolumn{2}{c |}{Formation} 
& \citep{Ligterink2018,Molpeceres2021,Tsai2022,Sanz-Novo2025,Perrero2025} \\ \cline{2-4}
&  \multicolumn{2}{c |}{Isomerisation} 
& \citep{Tsai2022,Molpeceres2022} \\  \cline{2-4}
\hline
\multirow{2}{*}{\bf Stereoselective kinetics in gas}
&  \multicolumn{2}{c |}{Formation} 
& see e.g., \citep{Cole2012,Neill2011,Ye2023,Mallo2025} \\ \cline{2-4}
&  \multicolumn{2}{c |}{Isomerisation (SAB)} 
& \citep{GarciadelaConcepcion2023,Neill2012,Sanz-Novo2025} \\ \cline{2-4}
\hline
\multirow{4}{*}{\bf Desorption-induced isomerisation (IUD)}
& \multirow{3}{*}{Non-thermal} & Chemical
& \citep{Molpeceres2025} \\ \cline{3-4}
& & Shocks 
& this work, \citep{Perrero2025} \\ \cline{3-4}
& & Cosmic rays 
& this work \\ \cline{2-4}
& \multicolumn{2}{|c |}{ Thermal desorption }
& this work \\ \cline{2-4}
\hline
\multicolumn{3}{|l |}{\bf Photoisomerisation }
& \citep{Cuadrado2016} \\ \cline{1-4}
\multicolumn{3}{|l |}{\bf Thermal isomerisation }
& \citep{Rivilla2017} \\ \cline{1-4}
\multicolumn{3}{|l |}{\bf Isomerisation assisted by tunneling }
& \citep{GarciadelaConcepcion2021,GarciadelaConcepcion2022} \\ \cline{1-4}
\multicolumn{3}{|l |}{\bf Stereoselective destruction (RDP) }
& \citep{Shingledecker2020} \\ \cline{1-4}
\hline
\end{tabularx}
\end{table*}

We show in Figure \ref{fig:OSR_deltaE} the expected behaviour of the OSR according to Equation \ref{eq-OSR} (considering $g=1$), for different values of $T_\text{kin}$, covering the range spanned by the astronomical sources where the isomers have been detected, from 7.5\K to 300\K (Table \ref{tab:OSR_sources_list}). Superimposed to the thermodynamic expectations, we have added the values of the OSR derived in the ISM towards the different regions, using the values of $\Delta E$ for each stereoisomeric pair from Table \ref{tab:stereoisomer_pairs}. To directly compare with the thermodynamic curves (which considers $g=1$), the values of the OSR for \ce{CH3CH2OH}, \ce{CH3CH2OCHO}, \ce{CH2CHCH2CN}, \ce{CH3CH2CH2CN} and \ce{(CH3)2CHOH} have been corrected by multiplying by 1/$g$=1/2, 1/2, 1/2, 2 and 2, respectively. 

%

Figure \ref{fig:OSR_deltaE} shows that all the isomers with $\Delta E \le$ 600\K detected in hot sources ($>$100 K) nicely follow the thermodynamic ratio. 
These pairs comprise conformational isomers of the following molecules: $n$-propanol, ethanol, $n$-propyl cyanide, ethyl formate, $i$-propanol, ethylene glycol, thioformic acid and vinyl alcohol; and the geometric isomers cyanomethanimine and ethanimine.
The $\Delta E$ of these species comes from experimental measures for six of them, with uncertainties of $\sim$5-30$\%$, and from  theoretical calculations for the remaining four.
The OSR values vary from $\sim$1 to $\sim$0.1, showing a trend of decreasing OSR with increasing $\Delta E$, as expected by thermodynamics. 
This highlights the fact that some stereoisomeric ratios in the ISM can offer stringent standards for energy accuracy, which indeed improve the   
inherent uncertainties of the state-of-the-art theoretical methods described above.
The comparison with observations shows that
small differences in $\Delta E$ seem to play a role in shaping the OSR values.
As a consequence, we propose here that the stereoisomeric ratios of these species can be used as a tool to estimate the kinetic temperature in hot astronomical sources.


Regarding the geometric isomers with $\Delta E \le$ 600\K (both imines), they fulfill the thermodynamic expectation despite the very high $E_{\rm iso}$ of $>$13000 K (not possible to overcome in the ISM). \citet{GarciadelaConcepcion2021} demonstrated that the thermodynamic ratio of these pairs is expected to be reached under insterstellar conditions through multidimensional ground-state quantum tunneling effects, 
unlike previously thought.\cite{Vazart2015,Balucani2018,Baiano2020}. Thanks to the tunneling, although the available energy is not enough to overcome the isomeric barrier, the system has a finite probability to pass through the barrier in both directions, from the lower-energy to the higher-energy isomer, and vice versa. In this context, the multidimensional character of the reaction path leads to significantly different effective rate constants in each direction. In this sense, the higher the energy difference between the isomers ($\Delta E$), the lower the rate of the isomerisation from the lower-energy to the higher-energy isomer, making more difficult to reach of the thermodynamic equilibrium.

%




An analogous tunneling-driven mechanism has been demonstrated to work efficiently for HCOSH ($\Delta E$=342 K) at high kinetic temperatures,\cite{GarciadelaConcepcion2022} which would explain the tentative detection of the higher-energy isomer towards the G31.41+0.31 hot core,\cite{GarciadelaConcepcion2022} although it is not efficient at low temperatures. 
Therefore, we can hypothesize that a similar mechanism might also work for \ce{CH3CH2OH}, \ce{(CH3)2CHOH}, and \ce{(CH2OH)2}, for which the isomerisation only requires a change in the position of an hydrogen atom (see Figure \ref{fig-3D}). They exhibit values of $\Delta E$ of 57 K, 120 K, and 300 K, respectively (Table \ref{tab:stereoisomer_pairs}), below the $\Delta E$ of the isomeric pairs for which tunneling has already been proven. Meanwhile, for \ce{H2CCHOH}, it remains unclear whether a similar process could occur, since the $\Delta E$ is somewhat larger (550 K). To our knowledge these tunneling processes proposed here have not yet been theoretically studied under ISM conditions, although they certainly merit attention. 

In contrast, for other pairs such as the $Gg'$/$Ag$ ($sc-sc'$/$ap-sc$) and $Aa$/$Ga$ conformers ($ap-ap$/$sc-ap$) of $n$-propanol, the $gauche$/$anti$ ($sc$/$ap$) conformers of $n$-propyl cyanide, and the $anti$/$gauche$ ($ap$/$sc$) conformers of ethyl formate, interconversion among the members of each pair is in principle not feasible through quantum tunneling, as it involves the displacement of a heavy -CH$_2$CH$_3$ fragment. Nevertheless, quantum tunneling may efficiently interconvert the $Ga$ and $Gg'$ conformers, and the $Aa$ and $Ag$ conformers, respectively, for which the $\Delta E$ are only 24 K and 73 K\cite{Jimenez-Serra2022}, respectively, because these interconversions imply only the displacement of an hydrogen atom. In fact, the OSR limit for $Gg'/Ga$ found in G+0.693 ($>$0.1\cite{Jimenez-Serra2022}) is consistent with the expected thermodynamic ratio of $\sim$0.6 at a $T_\text{kin}$=140 K, while in Sgr B2(N), the non-detections of the $Ga$ and $Aa$ forms are consistent with the thermodynamic ratio at 225 K\cite{Belloche2022} ($Gg'$/$Ga$ $\sim$ 0.72 and $Ag$/$Aa$ $\sim$ 0.9).

Regarding ethyl formate, since the $E_{\rm iso}$ is relatively low (348 K), an additional direct thermal isomerisation in gas phase cannot be ruled out in hot sources, 
and it may also contribute to the observed relative population of the conformers. For $n$-propanol and $n$-propyl cyanide, by contrast, all the observed conformers are populated in accordance with the expected thermodynamic ratio, including species connected by high $E$$_{iso}$ (1500-1750 K), suggesting that additional mechanisms are required to account for their observed OSR.
The only OSR derived for a cold source with $\Delta E \leq$ 600\K, which corresponds to \ce{CH2CHCH2CN}, significantly deviates from the expected thermodynamic ratio. The OSR is 1.14$\pm$0.16, several orders of magnitude higher than the thermodynamic expected value in a cold source such as TMC-1, where $T_\text{kin}\sim$ 9\K (Table \ref{tab:OSR_sources_list}). 
%
%
Similarly, the four stereosisomeric pairs with  $\Delta E >$ 600\K (up to $\sim$2700\K) $-$ \ce{CH3NHCHO}, \ce{CH3CHCHCN}, \ce{HCOOH} and \ce{CH3OCHO} $-$, also clearly deviate from the thermodynamic expectations (Figure \ref{fig:OSR_deltaE}), regardless the $T_\text{kin}$ of the astronomical source. 
In all cases the deviations are of several orders of magnitude.
It is clear in these cases that whether MEP would govern the isomeric population, the abundance of the higher energy steroisomers would be negligible, preventing their interstellar detections.  
As a consequence, alternative mechanisms need to be invoked. These can include chemical kinetic processes such as stereoselective formation\cite{Ye2023} and isomerisation, both on grain surfaces and in gas phase, and also desorption-induced isomerisation (Table \ref{tab:mechanisms}). This later mechanism, 
recently named as isomerisation-upon-desorption (IUD)\cite{Molpeceres2025}, is based on the hypothesis that an energy excess remain available on the molecules during (or just after) desorption. 
This process can be thermal, due to protostellar heating, but it might also occur  during energetic non-thermal events,  
such as cosmic-ray impacts and shocks\cite{Perrero2025}, in which the temperature can transiently reach peak values from several tens up to thousands of K\cite{Reboussin2014,Jimenez-Serra2008}.
%
In this context, once the molecules reach the gas phase, further interconversion is expected to be inefficient, and the conformer population would remain essentially frozen. This implies that the OSR would reflect the physical conditions at the time of desorption, rather than those currently present in the gas, and consequently the OSR can significantly deviate from the thermodynamic ratio assuming $T_{\rm kin}$ shown in Figure \ref{fig:OSR_deltaE}.







In the case of \ce{CH3NHCHO}, the OSR recently found towards the molecular cloud G+0.693-0.027 by Zeng et al.\cite{Zeng2025} is $2.9 \pm 0.6$, which is a factor of $\sim$120 higher than that expected according to thermodynamics at the $T_\mathrm{kin}$ of the source of $\sim$140 K (Table \ref{tab:OSR_sources_list}).
As discussed by Zeng et al.\cite{Zeng2025}, two possible mechanisms might explain the OSR. 
One option, based on the matrix experiments by Tsai et al.\cite{Tsai2022}, is a barrierless H abstraction of the lower energy $trans$ ($sp$) conformer, leading to the formation of the $trans$ ($sp$) \ce{CONHCH3} radical, followed by a second H abstraction that produces \ce{CH3NCO}. Once \ce{CH3NCO} is formed, hydrogenation yields the $cis$ ($ap$) \ce{CONHCH3} radical, which is subsequently hydrogenated to produce $cis$ ($ap$) \ce{CH3NHCHO}. Interestingly, the hydrogenation of \ce{CH3NCO} in laboratory ice experiments starting from \ce{CH4}:HNCO ice mixtures appears also to selectively form $cis$ ($sp$) \ce{CH3NHCHO}, since $trans$ ($ap$) \ce{CH3NHCHO} is not detected. \citep{Ligterink2018}

A possible alternative would be the gas-phase route proposed by Mirzanejad and Varganov,\cite{Mirzanejad&Varganov2025} which involves a spin-forbidden reaction between \ce{CH2} and formamide (\ce{NH2CHO}). This barrierless process proceeds through an intermediate that can occur as two chiral isomers, which upon H abstraction can eventually yield both $trans$ ($sp$) and $cis$ ($ap$) conformers of \ce{CH3NHCHO}. The formation rate constants of both isomers are expected to be similar, implying that they would be produced in comparable abundances. Nevertheless, the OSR in G+0.693-0.027 is $\sim$3:1. Hence, although this pathway alone not fully explains the measured value, it can certainly contribute to significantly increase the abundance of the higher-energy conformer.
%

For HCOOH, the energy difference between the two conformers is high, 2033 K (Table \ref{tab:stereoisomer_pairs}), and consequently the higher-energy $cis$ ($ap$) conformer should not be observed in space if they were populated according to thermodynamic equilibrium. However, it was first detected by Cuadrado et al.\cite{Cuadrado2016} in the Orion bar, which is a photo-dominated region illuminated by the massive Trapezium stars. The OSR is high, 0.35$\pm$0.13. These authors, based on $ab-initio$ calculations, explained the high abundance of $cis$ HCOOH in this environment via photo-isomerisation. The ultraviolet stellar photons absorbed by the conformers can radiatively excite them to electronic states above the interconversion barrier. Afterwards, subsequent fluorescent decay leaves the molecule in a different conformer, completing the photoswitching. This mechanism succeeded to explain the OSR in a photodominated region, but is not able to explain the presence of the higher-energy conformer in other environments where it has been found, such as the cold regions B5, L483 and TMC-1\cite{Taquet2017,Agundez2019,Molpeceres2025}, and the shock-dominated molecular cloud G+0.693-0.027.\cite{Sanz-Novo2023}
Garc\'ia de la Concepci\'on\cite{GarciadelaConcepcion2022} performed a theoretical study to evaluate whether quantum tunneling might account for the detection of the $cis$ ($ap$) conformer in these other regions. They found that the $cis$/$trans$ ($sp/ap$) ratio cannot be explained by isomerisation due to tunneling, and thus it should be determined through other chemical processes. In this context, two possible explanations have been proposed. First, a sequential acid-base (SAB) mechanism\cite{GarciadelaConcepcion2023}, which comprises a cyclic process of destruction and backward formation involving HCOOH and very abundant proton donors (HCO$^+$) and bases (NH$_3$). 
And second, the isomerization upon desorption recently suggested in an astrochemical modeling study by Molpeceres et al.\cite{Molpeceres2025}.
Additionally, it has been proven that a stereoisomeric excess of the most stable $trans$-HCOOH ($sp$-HCOOH) can be produced
via conformational isomerisation on grain surfaces,\cite{Molpeceres2022} i.e., through a two step hydrogen interconversion (H-abstraction + H-addition).


In the case of \ce{CH3OCHO}, the higher-energy $trans$ conformer lies 2667 K above the $cis$ ($sp$) form (Table \ref{tab:stereoisomer_pairs}). 
\cite{Faure2014, Sanz-Novo2025} Under thermal conditions, the expected $trans$/$cis$ ($ap/sp$) ratio would be $\sim$ 10$^{-13}$:1 at 100\K, implying that $trans$ ($ap$) \ce{CH3OCHO} should not be observable in the ISM. However, the OSRs span values between 0.029 and 0.08 (see Table \ref{tab:stereoisomer_pairs}), several orders of magnitude higher than the thermodynamic prediction. 
For this species, three stereoselective chemical pathways possibilities have been invoked to explain the OSR: 

(a) Stereoselective or preferential formation processes on grain surfaces (the \ce{CH3O} + \ce{HCO} route) can qualitatively account for the observed $cis$/$trans$ ($sp/ap$) abundance ratio, \cite{Sanz-Novo2025} suggesting that the OSR may be inherited from grain chemistry. A similar mechanism has also been suggested for HCOSH, and although its OSR follows the thermodynamic ratio (Figure \ref{fig:OSR_deltaE}), the formation of the low-energy $trans$ ($sp$) HCOSH has been found to be highly efficient through the hydrogenation of OCS on ices.\cite{Molpeceres2021}

(b) The gas-phase ion--molecule reaction from protonated methanol, \ce{CH3OH2+ + HCOOH -> HCOHOCH3+ + H2O}, which is exothermic and barrierless.\cite{Neill2011, Cole2012} This route involves two distinct transition-state geometries that produce different conformations of the protonated \ce{CH3OCHO}, depending on the $cis$ or $trans$ arrangement of the C-O-C-O dihedral angle. Notably, the formation of $trans$ ($ap$) \ce{CH3OCHO} is barrierless, whereas the production of the low-energy $cis$ conformer requires a net activation barrier of $10 \,\text{kJ mol}^{-1}$ (1200 K). Consequently, this route appears to be stereoselective, preferentially yielding $trans$ ($ap$) \ce{CH3OCHO} via dissociative recombination.

(c) The indirect isomerisation pathway involving protonation of \ce{CH3OCHO}, after which \ce{CH3OCHOH+} may undergo dissociative recombination. \cite{Neill2012,Sanz-Novo2025} Both steps are highly exothermic,\cite{Hunter1998} and the released energy may exceed the high $E_{\rm iso}$ = 6945\K (Table \ref{tab:stereoisomer_pairs}). Given similar barriers in the protonated form, isomerisation can effectively occur during the relaxation phase of either step potentially leading to a nonthermal OSR and enhancing the formation of $trans$ ($ap$) \ce{CH3OCHO}. This route is essentially analogous to the aforementioned SAB mechanism used to explain formic acid (\ce{HCOOH}) isomerism,\cite{GarciadelaConcepcion2023} suggesting that a common mechanism may underlie selective isomerism in several interstellar molecules.


Lastly, the two geometric isomers of crotononitrile ($trans$ and $cis$, or $Z$ and $E$ \ce{CH3CHCHCN}) were identified towards TMC-1 by Cernicharo et al.\cite{Cernicharo2022}. These stereoisomers are separated by 1132\K, with the $trans$ ($Z$) form being the global minimum in energy. Interestingly and despite the high $E_{\rm iso}$ of 25815 K\cite{Butler1963} (Table \ref{tab:stereoisomer_pairs}), the $cis$ ($E$) conformer  was found to be more abundant, obtaining a OSR of 2.6 $\pm$ 0.5, several orders of magnitude larger than that expected assuming thermodynamic equilibrium. While the formation of these species is still not fully understood, the temperature dependence of the $cis$-$trans$ ($E/Z$) ratio of \ce{CH3CHCHCN} has been previously explored in the gas phase.\cite{Butler1963}. These authors found a steady value for the $cis$-$trans$ ($E/Z$) ratio of $\sim$2 upon pyrolysis of pure $E$-\ce{CH3CHCHCN}, consistent with the value derived in the ISM. 
Recently, Mallo et al.\cite{Mallo2025} have investigated the C$_3$H$_6$ + CN gas-phase reaction and reports isomer-specific branching ratios for the formation of $E$ ($cis$) and $Z$ ($trans$) crotononitrile. While their results demonstrate that gas-phase chemistry can produce both stereoisomers, the predicted branching ratio ($Z$/$E$ $\sim$ 0.75) differs significantly from the observed OSR ($\sim$ 2.6; Table \ref{tab:stereoisomer_pairs}). This indicates that stereoselective gas-phase formation alone is not sufficient to account for the observations, and suggests that additional processes, such as distinct destruction routes, or perhaps ion-molecule reactions enabling isomerization, must contribute.


Besides stereoselective chemical formation routes, the OSR can also be shaped by stereoselective destruction routes. In this sense, Shingledecker et al.\cite{Shingledecker2020} proposed the relative dipole principle (RDP) as a possible “rule of thumb”. According to this idea, when the chemistry of a family of isomers is broadly similar, the relative abundances of different species should follow proportionally the inverse trend of their permanent dipole moments, because more polar species are expected to experience faster destruction by hydrogen atoms. However, while the RDP might contribute in shaping some of the OSRs by enhancing the destruction of highly polar stereoisomers under specific conditions, as observed for $Z$- and $E$-HNCHCN, \cite{Shingledecker2020} it certainly cannot account for all the OSRs studied in this work. As mentioned before, the dipole moments of the higher-energy stereosiomers are usually larger ($\mu_{\rm high}$/$\mu_{\rm low}\sim$0.9$-$3.1; see Table \ref{tab:stereoisomer_pairs}), but this factor is significantly smaller than those produced by thermodynamics, which depends exponentially on the $\Delta E$ (see curves in Figure \ref{fig:OSR_deltaE}), or by the deviations from equilibrium pointed out by the observations.



Taken altogether, this review of interstellar stereoisomerism underscores the relevance for astrochemistry of higher-energy stereoisomers, which have been traditionally overlooked due to the prevailing assumption that they should not be present in the ISM. We stress the importance of exploring the full isomeric panorama within a given family from both an experimental (i.e., spectroscopic and kinetic investigations) and theoretical point of view. Regarding high-resolution rotational studies, spectroscopic data at high frequencies (i.e., millimeter- and submillimeter-wavelengths) are typically available only for the global minimum in energy, while higher-energy stereoisomers often remain unexplored. This limitation arises because conventional frequency-modulated millimeter-wave experiments are usually performed at room temperature, conditions under which the population of higher-energy stereoisomers is strongly disfavored. Nevertheless, we stress that the laboratory characterization of the rotational spectra of these species is essential, regardless of their relative energy or isomerization barriers, as such data are critical to guide future radioastronomical searches and detections. 

What is more, to date, quantum chemical computations on the reactivity and formation mechanisms of interstellar systems often neglects the full isomeric landscape. Incorporating all isomers, along with their specific formation pathways and interconversion mechanisms, into astrochemical models — many of which currently remain largely insensitive to stereoisomerism — represents a critical step toward elucidating their formation routes and explaining the OSRs. This will be crucial not only for explaining the OSR, but also for predicting the presence of currently undetected species, ultimately providing a more comprehensive picture of the chemical complexity of the ISM.

\section{\rm \bf  \large  Detectability of new interstellar sources through identification of higher-energy stereoisomers}
\label{sec-detectability}

The detectability of a molecular species through rotational spectroscopy depends on two main factors: the molecular abundance and the strength of the dipole moment ($\mu$).
As discussed above, the OSRs for several of the interstellar stereoisomers are close to 1 or even above, namely, the abundances of the higher-energy isomers are of the order of those of the most stable steroisomers. Considering this, the higher-energy steroisomers might be a better option for detecting new interstellar species, whether their dipole moment are significantly larger than that of their isomers.
%
This is because the larger the dipole moment the brighter the line intensities (following $\mu^2$), and thus the easier to detect the molecule.
Table \ref{tab:stereoisomer_pairs} lists the total dipole moments ($\mu$) of the different stereoisomeric pairs (see the references there). The dipole moments of the higher-energy stereoisomers are in general larger than those of the lower-energy counterparts, since the latter tend to adopt conformations in which dipole moments cancel more efficiently, thereby minimizing electrostatic repulsion. As a result, $\mu_{\rm higher}$/$\mu_{\rm lower}$ ranges from $\sim$1.0$-$3.1 in all cases, with the only exception of crotononitrile, in which it is in any case 0.9, close to unity. 
Therefore, the overall higher dipole moments of the higher-energy isomers favours their detectability, especially on those cases in which the difference is higher ($\mu_{\rm higher}$/$\mu_{\rm lower}\sim$2.6$-$3.1), which are \ce{HCOOH}, \ce{CH3OCHO} and \ce{HNCHCN}. 
An extreme case in this regard is carbonic acid, HOCOOH, for which only the higher energy $cis$–$trans$ (or $sp$–$ap$) conformer has been detected in the ISM, \cite{Sanz-Novo2023} while the lower-energy $cis$–$cis$ ($sp$–$sp$) conformer remains undetected due to its much lower dipole moment (approximately fifteen times smaller).\cite{Mori2011}

Following this fact, we propose a novel but powerful 
method to detect new interstellar species: in some cases the presence of a molecule can be confirmed by the identification of its higher-energy stereoisomer (which usually has higher dipole moment), especially when the most stable species is observationally inaccessible  due to its low or zero dipole moment. 
The detection of the higher-energy conformer of HOCOOH has already demonstrated the viability of such approach, and the increasing number of detections of other higher-energy stereoisomers (Fig. \ref{fig-timeline}), makes it a promising strategy. 
We note that this provide a direct proof of the presence of the molecule, in contrast with other methods that relies on the indirect proof of non-polar species through the detection of protonated species, such as \ce{N2}\cite{Linke1983}, \ce{CO2}\cite{Thaddeus1981}, NCCN\cite{Agundez2015,Rivilla2019}, or \ce{NC4N}\cite{Agundez2023_Dicyanopolyynes}.

We propose here below some promising stereoisomers for new interstellar detections:


$\bullet$ $sp$-glyoxal (OCHCHO): glyoxal is a pivotal species in astrochemsitry, in particular it is thought that it plays an essential role in the formation of more complex species on the icy grain mantles\cite{Simons2020,Leroux2021,Wang2024}. While the lower energy stereoisomer, $ap$, has zero dipole moment, its higher-energy counterpart, $sp$ (or $cis$) has a large dipole moment of $\mu$=3.4 D\cite{Hubner1997}. The energy difference between isomers is high, $\Delta E$ = 2237$\pm$69 K\cite{Hubner1997}, so the expected population of the $sp$ would be very low if they follow the thermodynamic expectation. However, this energy difference is very similar to other higher-energy stereoisomers already detected in the ISM that we have discussed in this review, such as \ce{HCOOH} or \ce{CH3OCHO} (Table \ref{tab:mechanisms}). Therefore, if some of the mechanisms discussed here that can enhance the abundance of the higher-energy species is at play also for glyoxal, the interestellar detection might be able through the identification of the $sp$ stereoisomer, given that its rotational spectroscopy has already been measured\cite{Hubner1997}.

$\bullet$ $sp$-oxalic acid (\ce{(COOH)2}): oxalic acid is one of the most abundant dicarboxylic acids identified in carbonaceous condrites\cite{Peltzer1984}, and its formation in laboratory experiments of acetic acid ices exposed to UV radiation through the direct
recombination of two COOH radicals has been proposed\cite{delBurgo2024}. The recent detection of the first interstellar species with three oxygen atoms, carbonic acid\cite{Sanz-Novo2023}, paves the way to the detection of species with even more number of oxygen atoms, in particular dicarboxylic acids, of which oxalic acid is the simplest representative. 
Similarly to glyoxal, while the lower-energy stereoisomer of oxalic acid $trans$-$trans$-$trans$ ($ap$-$ap$-$ap$) has zero dipole moment, the higher-energy $cis$-$trans$-$trans$ ($sp$-$ap$-$ap$) isomer of oxalic acid has a high dipole moment of $\mu$=3.073 D
\cite{Godfrey2000}, and it rotational spectroscopy has been measured in the laboratory\cite{Godfrey2000}, allowing its interstellar search. 

$\bullet$ Higher-energy stereoisomers of amino acids: despite amino acids have been identified in
chondritic meteorites\cite{Cronin1983,Burton2012}, comets\cite{Altwegg2016}, and asteroids\cite{Potiszil2023,Glavin2025}, their presence in the ISM still needs to be confirmed, despite numerous attempts\cite{Combes1996,Ceccarelli_structure_2000,Snyder2005,Jones2007,Cunningham2007,Rivilla2023}. Following the strategy proposed here, the higher-energy stereoisomers of two of the simplest amino acids, glycine and alanine, can be good options for their first interstellar detections, even more than the respective lower-energy counterparts. This is because their dipole moments are significantly higher by factors of 5 and 3.3, respectively\citep{Lovas1995,Godfrey1993}. The expected line intensities of the molecular emission scales with $\mu^2$, so this can certainly counteract the thermodynamic factor, considering that the relative stereoisomeric energies are not very high: $E$ = 403 K\cite{Sanz-Novo2019,Lattelais2011} and $\Delta E$ = 350 K\cite{Blanco2004}, respectively. 








\section{\rm \bf \large $\blacksquare$  SUMMARY AND CONCLUSIONS}
\label{sec-conclusions}


The growing number of detections of higher-energy stereoisomers in the ISM demonstrates that stereoisomerism is a fundamental aspect of interstellar chemistry that directly contributes to the molecular complexity. In this work, we have conducted the first comprehensive overview
of interstellar stereoisomerism, compiling all stereoisomeric pairs detected so far in the ISM and evaluating their observed abundance ratios in the context of their energetics, interconversion barriers, and formation/destruction pathways. The current sample of stereoisomers, although still limited, already reveals interesting trends. The main conclusions of this study can be summarized as follows:

\begin{itemize}

\item We propose a homogeneous and updated nomenclature for all stereoisomeric pairs studied, based on modern IUPAC recommendations, which is provided alongside the traditionally employed descriptors to enable direct comparison.

\item The total number of stereoisomeric pairs detected so far in the ISM are 16, of which 13 are conformational and 3 are geometric.
They are molecules with 5 to 12 atoms, and include carbon-, nitrogen- and sulfur-bearing species. The energy difference between the members of each isomeric pair ($\Delta E$) covers a wide range from $\sim$10\K to 2667\K. These stereoisomers have been detected towards different types of astronomical objects, which include dark molecular clouds, a photodissociation region, hot cores/corinos, and shocked-dominated regions, which span a wide range of gas kinetic temperatures ($T_{\rm kin}\sim$7.5$-$300 K).

\item The observed stereoisomeric ratios (OSR), defined as the column density ratio of the higher-energy isomer divided by that of the lower-energy isomer, span values from 0.009 to 4, being most of them $\leq$ 1, with a few exceptions in which the higher-energy isomer is more abundant.

\item Thermodynamic equilibrium, in which the so-called Minimum Energy Principe (MEP) is based, is not generally sufficient to explain interstellar stereoisomeric ratios. Although stereoisomers with small energy separations ($\Delta E \lesssim 600$ K) observed in hot environments ($T_\text{kin}>$100\K) 
follow nicely the thermodynamic expectations, many other systems show clear deviations: those detected in cold dark clouds with $T_{\rm kin}\sim$ 10 K, and those detected in hot regions with $\Delta E >$ 600 K. Several higher-energy stereoisomers are detected with abundances that exceed equilibrium predictions by several orders of magnitude, ruling out thermodynamics as the controlling factor.
    
\item Higher-energy stereoisomers are a widespread component of interstellar chemistry. The detection of stereoisomers with $\Delta E$ values exceeding 1000–2000 K demonstrates that relative energetic instabilities alone does not prevent their presence in the ISM. Their existence requires the action of alternative mechanisms such as stereoselective formation routes, quantum tunneling-driven interconversion mechanisms, photoisomerisation, or chemical rearrangement during the desorption process.
    
\item Stereoisomeric ratios provide direct constraints on chemical pathways. The observed OSRs encode information on the underlying chemistry and in many cases cannot be reproduced without invoking stereoselective formation and destruction mechanisms. 

\item We propose that the detection of higher-energy stereoisomers (which usually have higher dipole moments than their lower-energy counterparts) can be a powerful method to reveal the presence of new molecules in the ISM, especially when the lower-energy stereoisomers have very low or zero dipole moments.
    
\item Laboratory spectroscopy of higher-energy stereoisomers is urgently needed to allow their interstellar search. Many high-resolution rotational spectra are currently available only for the lowest-energy stereoisomers. However,  higher-energy stereoisomers are not only detectable but, in some cases, abundant. 

\item We have performed new DFT theoretical calculations of isomerization energy barriers for the target molecules lacking prior data. New quantum theoretical calculations devoted to the study different direct or indirect isomerisation processes, as well as new stereoselective formation and destruction routes, are needed to explain the isomeric ratios observed.  Particularly, a detailed study of the quantum tunneling interconversion for \ce{CH3CH2OH}, \ce{(CH3)2CHOH}, \ce{(CH2OH)2}, and \ce{H2CCHOH} certainly merit attention.
    
\item Astrochemical models must explicitly include stereochemistry. Current chemical networks largely neglect stereoisomerism and therefore cannot reproduce observed stereoisomeric ratios. Incorporating stereoselective chemistry, including isomer-specific reaction pathways and destruction processes, is essential to accurately model molecular abundances in the ISM.

\end{itemize}

In summary, interstellar stereoisomerism emerges as a powerful diagnostic of chemical processes in the ISM. Therefore, a coordinated effort combining laboratory spectroscopy, quantum chemical calculations, chemical kinetics, and astronomical observations will be crucial to unravel the origin of stereoisomeric selectivity in the ISM and to achieve a more complete understanding of molecular complexity in space.

\begin{acknowledgement}

We are very grateful for the careful and detailed review carried out by the five independent reviewers, whose comments and suggestions have significantly helped to clarify many aspects of this work.
V.M.R. acknowledges support from the grant PID2022-136814NB-I00 by the Spanish Ministry of Science, Innovation and Universities/State Agency of Research \newline MICIU/AEI/10.13039/501100011033 and by ERDF, UE;  the grant RYC2020-029387-I funded by MICIU/AEI/10.13039/501100011033 and by "ESF, Investing in your future", and from the Consejo Superior de Investigaciones Cient{\'i}ficas (CSIC) and the Centro de Astrobiolog{\'i}a (CAB) through the project 20225AT015 (Proyectos intramurales especiales del CSIC); and from the grant CNS2023-144464 funded by MICIU/AEI/10.13039/501100011033 and by “European Union NextGenerationEU/PRTR”. M.S.N. acknowledges a Juan de la Cierva Postdoctoral Fellow project JDC2022-048934-I, funded by MICIU/AEI/10.13039/501100011033 and by the European Union “NextGenerationEU/PRTR”. D.S.A. expresses his gratitude for the funds received from the Comunidad de Madrid through the Grant PIPF-2022/TEC-25475, for grant CNS2023-144464 funded by MICIU/AEI/10.13039/501100011033 and by “European Union NextGenerationEU/PRTR”, as well as for the financial support provided by the Consejo Superior de Investigaciones Cient{\'i}ficas (CSIC) and the Centro de Astrobiolog{\'i}a (CAB) through the project 20225AT015 (Proyectos intramurales especiales del CSIC).

\end{acknowledgement}

\section*{Conflict of interest}
The authors declare no competing financial interest.



\bibliography{biblio}

\end{document}